\documentclass[aps,prd,twocolumn,showpacs,superscriptaddress,nofootinbib,preprintnumbers]{revtex4-2}

\usepackage{cancel}
\usepackage{style}
\usepackage{enumitem}
\usepackage{rotate}
\usepackage{bbm}
\usepackage{bm}
\usepackage{soul}
\usepackage[font=small,labelfont=bf]{caption}
\usepackage{subfiles}

\newcommand{\beq}{\begin{equation}}
\newcommand{\eeq}{\end{equation}}

\newcommand{\Xiexp}{\delta_{\mathrm{max}}}
\newcommand{\UFN}{$U(1)_{\rm FN}\,$}
\newcommand{\Oone}{$\mathcal{O}(1)$}

\begin{document}

\preprint{\tt CERN-TH-2024-215} 
\preprint{\tt  FERMILAB-PUB-24-0938-T}

\vspace{3cm}

\title{Testing the Froggatt-Nielsen Mechanism with Lepton Flavor and Number Violating Processes}

\author{Claudia Cornella\,\orcidlink{0000-0002-2573-2123}}
\email{claudia.cornella@cern.ch}
\affiliation{Theoretical Physics Department, CERN, Geneva, Switzerland}

\author{David Curtin\,\orcidlink{0000-0003-0263-6195}}
\email{dcurtin@physics.utoronto.ca}
\affiliation{Department of Physics, University of Toronto, Canada}

\author{Gordan Krnjaic\,\orcidlink{0000-0001-7420-9577}}
\email{krnjaicg@fnal.gov}
\affiliation{Theoretical Physics Division, Fermi National Accelerator Laboratory, Batavia, IL, USA}
\affiliation{Department of Astronomy \& Astrophysics, University of Chicago, Chicago, IL USA}
\affiliation{Kavli Institute for Cosmological Physics, University of Chicago, Chicago, IL USA}

\author{Micah Mellors\,\orcidlink{0009-0001-7786-9703}}
\email{m.mellors@mail.utoronto.ca}
\affiliation{Department of Physics, University of Toronto, Canada}

\begin{abstract}
The Froggatt-Nielsen (FN) mechanism offers an elegant explanation for the observed masses and mixings  of Standard Model fermions. In this work, we systematically study FN models in the lepton sector, identifying a broad range of charge assignments (``textures") that naturally yield viable masses and mixings for various neutrino mass generation mechanisms. Using these textures, we consider higher-dimensional operators consistent with a FN origin and find that natural realizations predict distinct patterns in lepton flavor- and number-violating observables. 
For Dirac and Majorana neutrinos, FN-related correlations can lead to detectable rates of charged lepton flavor violation at next-generation low-energy experiments.
Majorana and type-I seesaw models predict measurable rates of neutrinoless double beta decay.  
Determination of inverted neutrino mass ordering would exclude the Dirac neutrino FN scenario.
Only a small minority of purely leptonic FN models predict detectable flavor violation at future muon colliders, though it is possible that a combined analysis with the quark sector will reveal motivated signals.
These findings highlight the power of the FN mechanism to link neutrino mass generation to testable leptonic observables, offering new pathways for the experimental exploration of lepton number and underscoring the importance of next-generation low-energy probes. 
\end{abstract}

\bigskip
\maketitle


\section{Introduction}
\label{sec:intro}

Standard Model (SM) fermions exhibit a broad range of masses and mixing angles with distinct patterns. While technically natural, this dramatic variation invites an explanation beyond the SM.  The Froggatt-Nielsen (FN) mechanism \cite{Froggatt:1978nt} provides an elegant and economical framework for explaining this structure through a spontaneously broken horizontal symmetry.  

In these models, fermions carry additional FN charge and Yukawa couplings are forbidden at tree level. At low energies, the new symmetry is spontaneously broken and  heavy fields  are  integrated out, yielding effective Yukawa couplings whose magnitude is exponentially sensitive to the FN charges of the corresponding fermions. Thus, order one differences in charge assignments generate large hierarchies in masses and mixing parameters. 

This mechanism has been studied in both the quark and lepton sector~\cite{Leurer:1992wg, Leurer:1993gy, Dudas:1996fe, Irges:1998ax,   Sato:2000ff, Sato:2000kj, Suematsu:2001sp, Dreiner:2003hw, Nir:2004my, Kamikado:2008jx, Plentinger:2008up, Buchmuller:2011tm, Krippendorf:2015kta, Bauer:2016rxs, Ema:2016ops, Calibbi:2016hwq, Nishiwaki:2016xyp, Vien:2019zhs, Feruglio:2019ybq, Berger:2019rfk, Bordone:2019uzc, Smolkovic:2019jow, Fedele:2020fvh, Nishimura:2020nre, Aloni:2021wzk, Ringe:2022rjx, Asadi:2023ucx, Qiu:2023igq, Cornella:2023zme, Nishimura:2024apb, Ibe:2024cvi}. The lepton sector, which is the focus of this work, introduces the additional challenge of integrating neutrinos into the framework. Unlike quarks, whose masses and mixings all arise from Dirac-like Yukawa couplings, the underlying mechanism for generating neutrino masses is currently unknown, leading to a greater variety of possible implementations.
Furthermore, the Pontecorvo Maki Nakagawa Sakata (PMNS) matrix does not exhibit any hierarchies and there are large experimental uncertainties on the parameters of this sector; indeed, the precise values of neutrino masses are currently unknown.  

In this work we address these challenges systematically to present for the first time a \textit{global} picture of the expected relative magnitudes of CLFV observables resulting from a large number of phenomenologically viable lepton FN textures. We also present the correlations between observables which could in principle be used to discriminate between textures. We achieve this through a two step procedure: \\
{\bf 1. Identify realistic natural FN textures:} 
    We first scan over a wide range of FN charge assignments (``textures") for the leptons, and identify hundreds of combinations that naturally yield viable lepton masses and mixings. In our treatment, all other parameters (e.g. the coefficients of FN-preserving operators) are chosen to be generic order-one numbers such that the 
    hierarchies in the lepton sector arise entirely from the charge assignments and corresponding FN spurion insertions. \\ 
    {\bf 2. Predict CLFV and $0\nu \beta\beta$ for these textures:}
    Although the SM does not predict observable levels of charged lepton flavor violation (CLFV) or neutrinoless double beta decay ($0\nu\beta\beta$), such processes can be greatly enhanced in the presence of higher-dimensional operators that arise from integrating out the FN sector. The relative importance of different observable processes is dictated by texture-specific selection rules for each model. %
    We therefore calculate the predicted rates of lepton violating processes for the identified realistic textures within the Standard Model Effective Theory (SMEFT) framework, which allows us to \emph{evaluate the experimental prospects of the leptonic FN mechanism as a whole.}

We consider neutrino mass generation by three mechanisms: Dirac, generic Majorana via a Weinberg operator, and type-I seesaw.\footnote{Note that while many type-I seesaw scenarios can be mapped directly to a Weinberg operator with a single Weinberg scale, as in the Majorana case, if the RH neutrinos have flavor charge the type-I seesaw would effectively generate a Weinberg operator where the suppression scale has flavor structure.}
Our analysis reveals how a variety of low-energy, high-energy, and cosmological probes can provide a direct window into the dynamics underlying lepton flavor.

\vspace{-0.5cm}


\section{Froggatt-Nielsen Neutrino Masses}
\label{sec:FNandneutrinos}
The FN mechanism introduces a \UFN \, symmetry  which is spontaneously broken at a UV scale $\Lambda$  by the vacuum expectation value (VEV) of a heavy scalar $\phi$.
$\epsilon \equiv \phi/\Lambda \ll 1 $
is the spurion associated with the breaking of \UFN. 
Without loss of generality, we take the FN charge of $\phi$ (SM Higgs) to be $X_\phi=1$ ($X_H=0$), and the VEV of $\phi$ to be in the positive real direction. The SM Yukawa couplings only arise through $\epsilon$ insertions after FN breaking.

In two-component fermion notation, the charged lepton Yukawa couplings take the form 
\beq\label{eq: charged lepton lagrangian}
    \mathcal{L}_Y \supset -c^\ell_{ij}  L_i H^\dagger \bar e_j  \epsilon^{|X_{L_i} + X_{\bar e_j} |} ,
\eeq
where $L_i$ are the left-handed (LH) lepton doublets, $e_j$ the right-handed (RH) charged lepton singlets, $H$ is the Higgs doublet,  $c^\ell$ a coupling matrix with \Oone\ entries, and $i,j=1,2,3$ label the fermion generations. 
The resulting Yukawa matrix is 
$
    (Y_\ell)_{ij} \equiv c^\ell_{ij}\epsilon^{n^\ell_{ij}} ,  n^\ell_{ij} \equiv |X_{L_i} + X_{\bar e_j}|,
$
naturally generating hierarchies when $\epsilon \ll 1$.  
Diagonalizing the Yukawas yields the charged lepton masses
$Y_\ell = U_\ell \hat Y_\ell W_{\ell}^\dagger\implies
\hat m_\ell =  \frac{v}{\sqrt{2}}\hat Y_\ell~
$
after electroweak symmetry breaking (EWSB), 
where $\hat Y_\ell$ is diagonal, $U_\ell$ and $\ W_\ell$ are unitary, and $v = 246$ GeV. 
To account for neutrino masses, we examine three generation mechanisms within the FN framework. 

{\bf Dirac -- } Here the SM is supplemented with right-handed (RH) neutrinos $N_i$, and neutrino masses only arise from the Yukawa couplings
\beq\label{eq: dirac lagrangian}
 ~~~ \mathcal{L}_D \supset c^\nu_{ij}  
 \epsilon^{n^\nu_{ij}} H
 {L}_{i}  N_j   ~~ , ~~
    n^\nu_{ij} \equiv | X_{L_i} + X_{N_{j}}|~, \!\!\!\!\!
\eeq
where $c^\nu$ is a matrix with order-one entries. 
In analogy with charged leptons, the neutrino Yukawa matrix can then be written as $(Y_\nu)_{ij} \equiv c^\nu_{ij} \epsilon^{n^\nu_{ij}}$ and yields
$Y_\nu = U_\nu \hat Y_\nu W_{\nu}^\dagger~\implies~
\hat m_\nu =  \frac{v}{\sqrt{2}}\hat Y_\nu,
$ where $\hat Y_\nu$ is the diagonal Yukawa matrix, $U_\nu$ and $W_\nu$ are unitary matrices, $\hat m_\nu$ is the diagonal mass matrix, and 
the PMNS  matrix can be written as $V \equiv U^{\dagger}_\ell U_\nu$. 

In this scenario, the FN sector preserves lepton number, and the smallness of neutrino masses is entirely due to the FN mechanism and large flavor charges for RH neutrinos. 

{\bf Majorana --} 
The Weinberg operator
\cite{Weinberg:1979sa}
\beq\label{eq: weinberg lagrangian}
 ~~~~~~~   \mathcal{L}_W \!\!\supset\! - \frac{c_{ij}^W \epsilon^{n^W_{ij}} }{\Lambda_W} (L_i H)(L_j H) ,
\eeq
generates neutrino masses,
where $n^W_{ij}\equiv \! |X_{Li}+X_{Lj}|,$ $c^W$ is a symmetric matrix with \Oone\ elements,
and $\Lambda_W$ is the effective scale at which the operator is generated, which generically differs from the FN scale $\Lambda$. After EWSB,  Majorana neutrino masses arise via
\beq
\label{eq: weinberg neutrino mass}
     \hat m_\nu =  
     U_\nu^T
     \left[ c^W \epsilon^{n^W}\brac{     v^2    }{\Lambda_W}  \right] U_\nu, 
\eeq
where $U_\nu$ is a unitary matrix, $\hat m_\nu$ is the diagonalized mass matrix, and $V = U_\ell^\dagger U_\nu$ is the PMNS matrix. 

{\bf Seesaw -- } 
We consider a type-I seesaw scenario~\cite{Minkowski:1977sc,Yanagida:1980xy,Gell-Mann:1979vob,Mohapatra:1979ia} where flavor breaking and lepton violation are governed by the FN mechanism. In this case, neutrinos receive both  Dirac and  Majorana mass terms: 
\be
\label{eq: type 1 seesaw lagrangian}
\mathcal{L}_{\rm SS} \supset -c^\nu_{ij} \epsilon^{n^\nu_{ij}} H  L_{i}  N_{j}  
    - c^M_{ij} \epsilon^{n^M_{ij}}  \frac{ M }{2}{N}_{i} N_{j}   ,
\ee
where $c^\nu$ and $c^M$ are matrices of order-one coefficients, $M$ is the Majorana mass scale of the RH neutrinos, and 
$ n^\nu_{ij} \equiv | X_{L_i} + X_{N_{j}}|, n^M_{ij} \equiv |X_{N_{i}}+X_{N_{j}}|$.
In the seesaw limit, the Dirac contribution is small and the diagonal neutrino mass matrix is 
\be\label{eq: seesaw neutrino mass}
\hspace{0.4cm}
  \hat  m_\nu \approx 
  \frac{v^2}{2 M}
  U_\nu^T  \! \left( c^\nu\epsilon^{n^{\nu}} \right) \!\!
    \left(c^M \epsilon^{n^M} \right)^{-1}
    \!
    \left( c^\nu \epsilon^{n^{\nu}} \right)^T \! U_\nu , ~~~~~
\ee
where $U_\nu$ is a unitary rotation matrix. 
It is natural to identify the FN scale with the Majorana mass scale ($M=\Lambda$).
There could also be additional explicit lepton number violation in the FN sector, but we find that this yields qualitatively similar results (see Appendix \ref{app: lepton number violation}).
Note that type-I seesaw models generate an effective Weinberg operator through RH neutrino exchange, which ultimately gives rise to active neutrino masses. However, as the RH neutrino masses are also set by the FN mechanism, the resulting effective $\Lambda_W$ carries strong flavor dependence, distinguishing this scenario from the minimal Majorana scenario introduced previously. 

Note that most explicit models of neutrino mass generation can be accommodated within these scenarios \cite{Magg:1980ut, Schechter:1980gr, Lazarides:1980nt, Mohapatra:1980yp, Wetterich:1981bx} (see Appendix \ref{app: uv completion majorana}). 

\section{Methods}
\label{sec:methods}

\noindent \textbf{Identifying Realistic Natural FN Textures --}
We begin by identifying viable FN textures that adequately reproduce the observed lepton masses and mixing parameters for a common value of $\epsilon$. To ensure that, for a given texture, all hierarchical structure arises only from $\epsilon$, we demand that  all other free parameters are order-one numbers.
In our analysis, these free parameters are encoded in the Lagrangian coefficients
$c^\ell_{ij},  c^\nu_{ij},  c^{W}_{ij},$ and  $c^{M}_{ij}$
from Eqs. \eqref{eq: charged lepton lagrangian}, \eqref{eq: dirac lagrangian}, \eqref{eq: weinberg lagrangian}, and \eqref{eq: type 1 seesaw lagrangian}, respectively.
We scan over a wide range of possible FN charges and determine the fraction of random ${\cal O}(1)$ values in these coefficients that reproduce all observed masses and mixings within some tolerance. 

Our scanning procedure avoids conducting direct fits to the experimental observables. Instead, we seek to exhaustively identify charge assignments that resemble our world over a majority of their natural parameter space, realizing the intended spirit of a \textit{natural} FN solution to the flavor problem.
Thus,  our predictions are derived for FN models which generate experimental predictions very close to their observed values, though the fit is not exact. However, we have confirmed that, for coefficients that approximately reproduce known results, introducing small post-hoc tweaks  readily accommodates all known observables exactly. Thus, our scanning strategy does not lose any essential generality by seeking out approximate fits to experimental data.

Specifically, we adopt and extend the Bayesian-inspired method of Ref.~\cite{Cornella:2023zme} to the lepton sector -- see Appendix \ref{app: numerics} for details. For each mass generation mechanism, we consider all textures with charges $|X|\leq 7$, and, for Dirac RH neutrinos, $|X|\leq 9$ as required to obtain viable models. For each texture, we generate random  coefficient matrices $c_{ij}$, with each $\log_{10}c_{ij}$ sampled from a normal distribution centered on zero with standard deviation $\sigma=0.3$, and phases sampled uniformly over $[0,2\pi]$. (Alternative choices of reasonable ``$\mathcal{O}(1)$'' priors do not meaningfully affect results.)

For each choice of coefficients, we compute observables
$
\mathcal{O} = \left\{
m_\ell ~,~
\Delta m^2_{ij}~,~|V_{ij}|~,~ \sum m_\nu\right\},
$
corresponding to the charged lepton masses, neutrino mass-squared differences, the absolute value of the PMNS matrix elements and the sum of neutrino masses.
The fractional deviation of each FN prediction $\mathcal{O}_{\rm FN}$ is
$
\delta_\mathcal{O} \sim \mathcal{O}_{\rm FN}/\mathcal{O}_{\rm exp},
$ where 
$\mathcal{O}_{\rm exp}$ is the  
experimentally measured value (see Appendix \ref{app: detailed methods SM fit}). We maximize over all observables to obtain the overall experimental deviation: $
\Xiexp \equiv \max_{\mathcal{O}} \, (\delta_\mathcal{O} ).
$
Next, we adjust $\epsilon$ and, where applicable, $\Lambda_{W}$ or $\Lambda$, to minimize $\Xiexp$.
For each texture, this process is repeated many times for many choices of order-one coefficients. Textures that naturally resemble our world will have $\Xiexp \sim \mathcal{O}(1)$ for a large fraction coefficient choices.
To compare textures, we define
$
\mathcal{F}_x \equiv \text{\% of 
coeff. choices for which} ~ \Xiexp\leq x ,
$
allowing for textures to be ranked by $\mathcal{F}_x$ for different $x$.

In the Dirac case, leptonic masses and mixings depend only on $\epsilon$, so the procedure outlined above leaves $\Lambda$ unconstrained. For the Majorana (type-I seesaw) scenario, neutrino masses and mixings explicitly depend on both $\epsilon$ and $\Lambda_W$ ($\Lambda$), constraining both when minimizing $\Xiexp$.

\begin{table*}
    \centering
    \hspace*{-6mm}
    \begin{tabular}{ccccc} 
    \textbf{Dirac} & \phantom{bla} &    \textbf{Majorana} & \phantom{bla} &
     \textbf{Type-I Seesaw}
    \\
   \scalebox{0.9}{ \begin{tabular}{cccccccccccccc}

    $L_1$ & $L_2$ & $L_3$ & $\bar e_1$ & $\bar e_2$ & $\bar e_3$ & $N_1$ & $N_2$ & $N_3$ & $\epsilon$ & NO \\
    \midrule
    6 & 5 & 5 & -3 & -2 & 0 & 9 & 8 & 8 & 0.10 & 96 \\
    3 & 3 & 3 & 2 & -1 & -6 & 9 & 9 & 8 & 0.07 & 99 \\
    3 & 3 & 3 & 2 & -5 & -6 & 9 & 9 & 8 & 0.07 & 99 \\
    7 & 7 & 6 & -4 & -2 & 0 & 9 & 9 & 9 & 0.14 & 99 \\
    7 & 7 & 6 & -4 & -3 & -1 & 9 & 7 & 7 & 0.11 & 99 \\
    3 & 3 & 3 & 2 & 0 & -5 & 9 & 9 & 8 & 0.07 & 99 \\
    3 & 3 & 3 & 2 & 0 & -1 & 9 & 9 & 8 & 0.07 & 99 \\
    6 & 5 & 5 & -3 & -2 & 0 & 9 & 7 & 7 & 0.08 & 97 \\
    7 & 3 & 3 & 2 & 0 & -5 & 9 & 9 & 9 & 0.08 & 93 \\
    6 & 6 & 6 & -4 & -3 & -1 & 9 & 6 & 5 & 0.07 & 99 \\
    \end{tabular}}

& &

 \scalebox{0.9}{\begin{tabular}{cccccccccccc}

$L_1$ & $L_2$ & $L_3$ & $\bar e_1$ & $\bar e_2$ & $\bar e_3$ & $\epsilon$ & $\log\Lambda$ & NO \\
\midrule
2 & 0 & -1 & 7 & 6 & 4 & 0.24 & 15 & 91 \\
5 & 5 & -2 & 7 & -2 & -3 & 0.08 & 12 & 3 \\
4 & 4 & 3 & 5 & 2 & 0 & 0.23 & 11 & 96 \\
7 & 6 & 5 & 7 & 3 & 0 & 0.39 & 11 & 97 \\
6 & 6 & 5 & 5 & 1 & -1 & 0.30 & 10 & 96 \\
7 & 7 & 6 & 2 & -1 & -3 & 0.23 & 7.6 & 96 \\
5 & 5 & 4 & 6 & 2 & 0 & 0.30 & 11 & 96 \\
7 & 7 & 6 & 4 & 0 & -2 & 0.30 & 9 & 96 \\
5 & 5 & -2 & 7 & -2 & -7 & 0.08 & 12 & 3 \\
1 & 1 & -1 & -7 & -5 & -4 & 0.18 & 15 & 2 \\
\end{tabular}}

& &

\scalebox{0.9}{\begin{tabular}{ccccccccccccccc}
    $L_1$ & $L_2$ & $L_3$ & $\bar e_1$ & $\bar e_2$ & $\bar e_3$ & $N_1$ & $N_2$ & $N_3$ & $\epsilon$ & $\log\Lambda$ & NO \\
    \midrule
    6 & 1 & -1 & 7 & 7 & 6 & 3 & 0 & -4 & 0.36 & 14 & 93 \\
    6 & 1 & -1 & 6 & 6 & 6 & 3 & 0 & -4 & 0.34 & 14 & 93 \\
    6 & 1 & -2 & 7 & 7 & 7 & 5 & 0 & -4 & 0.37 & 14 & 93 \\
    7 & 2 & -1 & 7 & 7 & 7 & 4 & 0 & -5 & 0.40 & 14 & 95 \\
    6 & 2 & -6 & 2 & 1 & 1 & 3 & 2 & -4 & 0.16 & 12 & 90 \\
    4 & 1 & -1 & 6 & 5 & 5 & 6 & 0 & -3 & 0.27 & 14 & 93 \\
    4 & 1 & -1 & 7 & 5 & 5 & 4 & 0 & -3 & 0.29 & 14 & 93 \\
    7 & 2 & -1 & 7 & 7 & 6 & 4 & 0 & -5 & 0.39 & 14 & 95 \\
    6 & 1 & -1 & 7 & 6 & 6 & 3 & 0 & -4 & 0.35 & 14 & 93 \\
    5 & 1 & -1 & 5 & 5 & 5 & 2 & 0 & -3 & 0.27 & 14 & 79 \\
    \end{tabular}}
    \end{tabular}
    \caption{
    Some of the most natural and realistic FN textures for  Dirac, Majorana, and type-I seesaw neutrinos, reproducing masses and mixings with a relative experimental deviation factor $\Xiexp <$ 5, 2, 1.35  for approximately 50\%, 2-5 \% and 0.03\% of  random $\mathcal{O}(1)$ coefficient choices, respectively.
    Each texture is specified by the FN charges of the LH lepton doublets ($X_{L_i}$), RH charged leptons ($X_{\bar e_i}$), and RH neutrinos ($X_{N_i}$). 
    For $\epsilon$ and $\log_{10}(\Lambda/\mathrm{GeV})$, we show texture-averages for coefficient choices with $\Xiexp < 2$.
    NO denotes  the percentage of coefficient choices that predict normal ordered (NO) neutrino masses. 
    }
    \label{tab: charges}
    \centering
\end{table*}

\medskip

\noindent \textbf{Predicting Lepton Violation in FN --}
Having identified the realistic and natural textures capable of reproducing leptonic masses and mixings, we next explore their implications for current and future experiments with a focus on lepton flavor violation. 
To remain agnostic about the UV completion of the FN mechanism, we adopt the SMEFT framework with the minimal assumption that $\epsilon$ is the only spurion of \UFN breaking. 
Assuming ${\cal O}(1)$ coefficients in the UV theory, higher-dimensional SMEFT operators are then suppressed  only by powers of  $\Lambda$ and insertions of $\epsilon$. 

The most relevant interactions for LFV processes are the four-lepton, dipole, and semi-leptonic operators.
The four-lepton terms are 
\beq
    \mathcal{O}_4 = \frac{c_{ijkl}}{\Lambda^2}  \left(\Bar{\psi}_i \psi_j \right) \left(\Bar{\psi}_k \psi_l \right) \epsilon^{n_{ijkl}},
\label{eq:scaling_4fermi}   
\eeq
where $c_{ijkl}$ are ${\cal O}(1)$ coefficients and we have defined
$
n_{ijkl} \equiv |X_{\psi_i} - X_{\psi_j} +X_{\psi_k} -X_{\psi_l}|.$
Note that we switched to four-component fermion notation to match SMEFT conventions, but we  write $n_{ijkl}$ in terms of the FN charges of the corresponding 2-component fermion fields.
For dipole operators, we have \beq 
    \mathcal{O}_d = \frac{c_{ij}}{\Lambda^2}(\bar L_i \sigma^{\mu\nu} e_j) H F_{\mu\nu}  \epsilon^{n_{ij}}
    ~,~
    n_{ij} \equiv | X_{L_i} + X_{\bar e_j} |,
    \label{eq:scaling_dipole}
\eeq
where $F_{\mu\nu}$ is the field strength tensor of an electroweak gauge boson. Rates for muon to electron conversion in atomic nuclei are also sensitive to semi-leptonic four-fermion operators of the form \Eq{eq:scaling_4fermi}, with one of the bilinears comprised of light quark fields ($u,d,s$). To be agnostic to the flavor structure of the quark sector, we include only the flavor diagonal quark operators.
Notably, the structure of these operators is such that textures with a non-zero Higgs charge can be mapped onto $X_H = 0$ textures without altering the phenomenology, as discussed in Appendix \ref{app: non trivially charged H}. 

In our analysis,  ${\cal O}(1)$ Wilson coefficients are generated in the FN basis at a high scale $\Lambda$ and subsequently rotated into the mass basis using the values of $\epsilon$ and the $U$ and $W$ matrices from Eqs. 
 \eqref{eq: weinberg neutrino mass}, \eqref{eq: seesaw neutrino mass}, 
built using the Lagrangian coefficients appropriate to each scenario.
We calculate low-energy observables, including two and three-body CLFV decays of muons and taus, as well as muon-to-electron conversion in nuclei, using the {\tt flavio} package \cite{Straub:2018kue}. Matching to the relevant energy scales is performed with {\tt wilson} \cite{Aebischer:2018bkb} (running was found to be a negligible effect). Based on experimental constraints, we derive lower bounds on $\Lambda$. Finally, using these bounds, we estimate the highest possible rates of $e\mu$, $e\tau$, and $\mu\tau$ production at future $e^+e^-$ and $\mu^+\mu^-$ colliders. 
For comparison, we also compute the predictions for a fully anarchic or \textit{null} texture, where all charges are set to zero, and the Wilson coefficients are assumed to be all \Oone\ at a fixed scale. 
Further details on our methodology can be found in Appendix \ref{app: detailed methods pheno}.

\section{Results}
\label{sec:results}

The top-ranked  textures for each mass generation mechanism are presented in Table~\ref{tab: charges} -- additional textures are provided in the ancillary files. For both the Dirac and Majorana cases, 
the hierarchical structure of the charged lepton masses primarily stems from the charges of the RH charged leptons, and this fact has significant phenomenological implications. In particular, the LH rotation matrices often feature large off-diagonal terms, 
leading to relatively uniform contributions to various LFV processes across different textures, ultimately making the FN mechanism more predictive. Purely leptonic FN models of any type only rarely generate observable flavor-violating signals at proposed muon colliders; see Appendix \ref{app: supplemental plots} for details

\begin{figure}[t]
    \centering
    \includegraphics[width=0.9\linewidth]{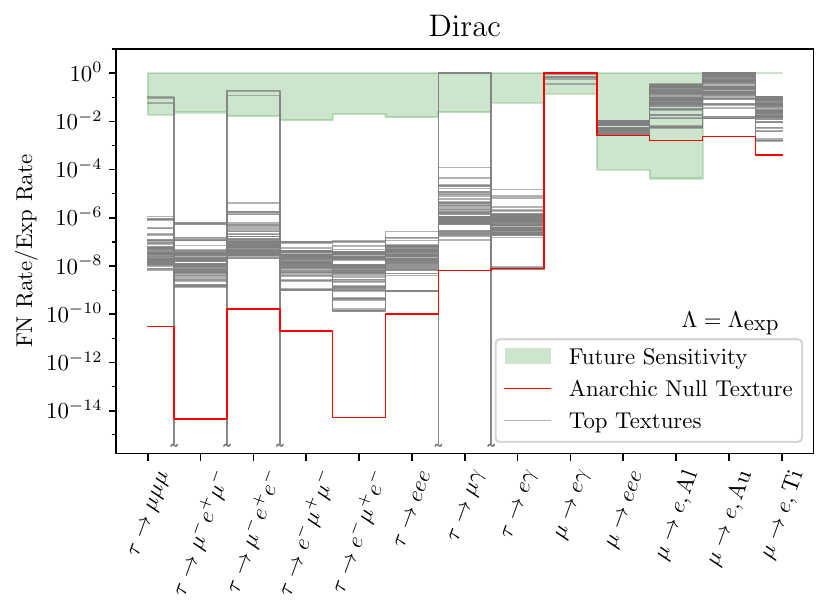} 
    \caption{ 
    Average predicted CLFV decay rates for the 100 most realistic natural Dirac FN textures (gray lines), relative to each observable's current constraint. 
    In each model, the flavor scale was chosen to saturate current experimental bounds at $\Lambda \sim 10^6$~GeV, thus fixing the other rates.  
    Green shading indicates the reach of proposed future low-energy CLFV experiments, and the flavor-anarchic null texture is shown as a red  line for comparison.    
    }
    \label{fig: Dirac results}
\end{figure}

\noindent \textbf{CLFV --}
\emph{Dirac:}
    In most Dirac FN models, the  FN scale is constrained to satisfy $\Lambda \gtrsim  10^{6}$ GeV by limits on $\mu \to e \gamma$~\cite{MEG:2016leq} or muon conversion in Gold~\cite{SINDRUMII:2006dvw} (see Appendix \ref{app: detailed methods pheno} for details). Setting $\Lambda$ to saturate its constraint within each individual model determines the highest possible rates for future CLFV signals. As shown in Fig.~\ref{fig: Dirac results}, 
Dirac FN textures can then predict measurable signals in LFV muon decays and $\mu$-$e$  conversion in atomic nuclei. The latter will be probed in new experiments with aluminum targets~\cite{Mu2e:2014fns} (though gold targets would provide the greatest discriminatory power between textures).
Furthermore, stringent constraints in the muon sector lead to suppressed LFV $\tau$ decays, with only a few  textures approaching detectability in $\tau$-related channels.
Overall, the predicted signal rates for FN Dirac scenarios typically exceed those of anarchic models by more than an order of magnitude, particularly for muonic processes, and offer a modest degree of discriminatory power between different textures.

\begin{figure}[t]
    \centering
    \includegraphics[width=0.9\columnwidth]{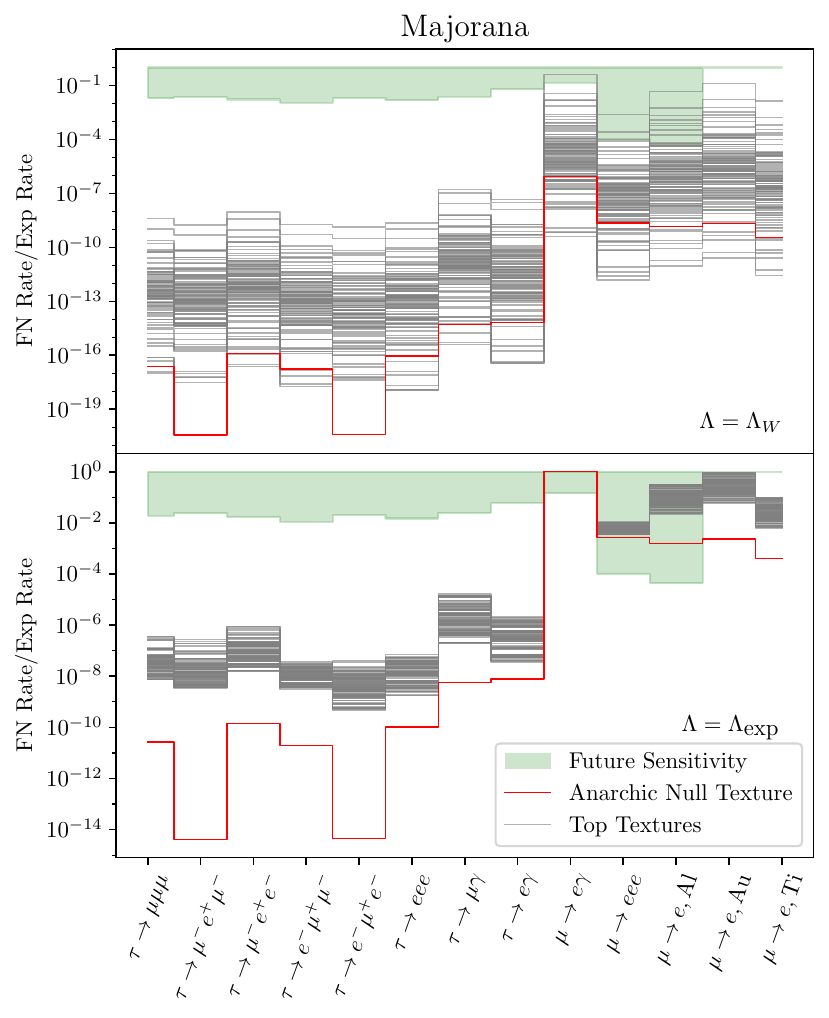}
    \vspace*{-3mm}
    \caption{
    Average predicted CLFV decay rates for the top most 100 realistic Majorana FN textures (gray lines), as in Fig.~\ref{fig: Dirac results}. 
    The top panel assumes that the scale of the Weinberg operator ($\Lambda_W$) coincides with the FN scale ($\Lambda$), fixing the predictions for CLFV signals from the imposed neutrino mass constraints. The \lq predicted\rq\ neutrino scale for the null texture is set to $10^{14}$ GeV.
    The bottom panel assumes $\Lambda_W \neq \Lambda$, such as in FN type-II seesaw scenarios, where $\Lambda$ is instead chosen to saturate its most restrictive current experimental bound of $\Lambda \sim 10^6$~GeV. }
    \label{fig: weinberg results}
\end{figure}

\emph{Majorana:} 
Results depend on whether the Weinberg operator scale $\Lambda_W$ coincides with the FN scale $\Lambda$. If $\Lambda_W=\Lambda$, the predictions for CLFV processes are fixed by neutrino masses, see Fig.~\ref{fig: weinberg results} (top). In this scenario, a non-observation in future experiments would exclude specific textures entirely. Unfortunately, most of the top Majorana textures correspond to scales $\Lambda_W \sim 10^{8}-10^{15}$ GeV, far beyond the reach of current or planned experiments. A handful of textures approach the observable region, exclusively in LFV muon experiments.
The assumption $\Lambda_W = \Lambda$ can be relaxed, as in FN type-II seesaw models. If we instead let $\Lambda$ saturate the most constraining current bounds, as in Fig.~\ref{fig: Dirac results}, it leads to the predictions shown in  Fig.~\ref{fig: weinberg results} (bottom). Majorana FN models then lead to similar, though slightly tighter, predictions in $\mu$-$e$  conversion experiments relative to the Dirac case.

\emph{Type-I Seesaw:}
These are the most challenging scenarios to probe experimentally. All observables are governed by the scale $\Lambda$, which is fixed by the scale of neutrino masses and  is too high to predict observable signals at future experiments for our best textures, with a small number of exceptions (see Appendix \ref{app: supplemental plots}). However, FN type-I seesaw scenarios can be probed more effectively via $0\nu\beta\beta$ decay experiments.

\begin{figure}
    \centering
    \hspace*{-5mm}
    \begin{tabular}{l}
\includegraphics[width=0.9\linewidth]{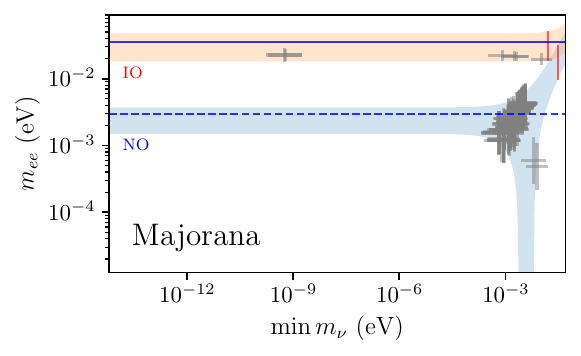}
    \\
\includegraphics[width=0.9 \linewidth]{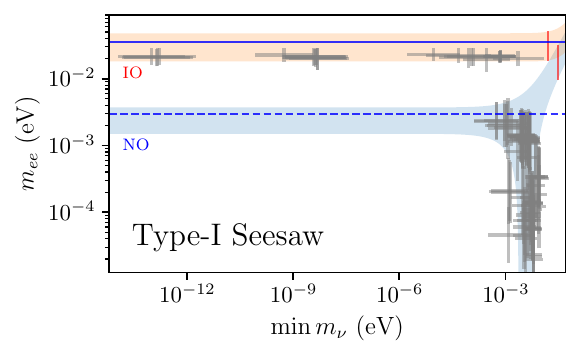}
    \vspace*{-0.5cm}
    \end{tabular}
    \caption{  Effective Majorana mass $m_{ee}$ versus the mass of the lightest neutrino for the top 100 FN textures in the Majorana  (top) and type-I seesaw (bottom) scenarios. 
    The yellow- and blue-shaded regions indicate the allowed ranges for IO and NO, respectively, based on the current measured values for $\Delta m_{32}^2$ and $\Delta m_{21}^2$. The solid blue line indicates the strongest exclusion from current $0\nu\beta\beta$ searches, from KamLAND-Zen  \cite{ParticleDataGroup:2022pth}, while the dashed blue line marks the reach of the future $0\nu\beta\beta$ experiments nEXO \cite{nEXO:2021ujk} and CUPID \cite{CUPID:2022wpt}. The vertical red lines indicate the limits on $\min m_\nu$ in the IO and NO scenarios from current cosmological limits on $\sum m_\nu$.  Error bars show the $1\sigma$ spread of predictions with $\Xiexp<2$. }
    \label{fig: seesaw 0nuBB}
\end{figure}

\begin{figure}[t]
    \hspace{-1cm}\includegraphics[width=0.9\linewidth]{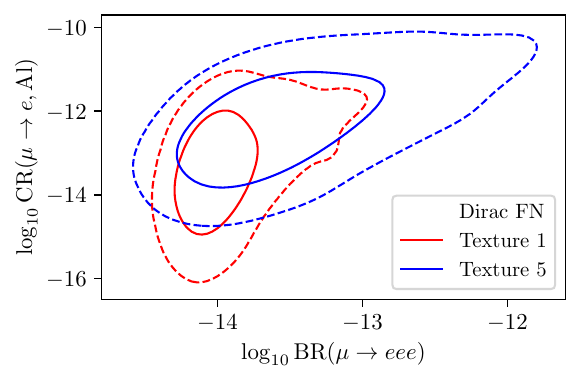}
    \vspace{-0.3cm}
    \caption{Comparison of the predicted branching ratio for $\mu \to 3 e$ and the conversion rate for $\mu$-$e$ conversion in Al for the first and the fifth Dirac FN textures in Table \ref{tab: charges}, assuming the FN scale $\Lambda$ saturates its current lower bound. The conversion rate is normalized to the muon capture rate in nuclei. Contours show areas containing 68\% (solid) and 98\% (dashed) of random coefficient choices with $\Xiexp<2$.  Thus, these textures might be distinguished despite similar average predictions.}
 \label{fig: correlations}
\end{figure}

Additional information can be gained from correlations among  lepton-violating observables, beyond their average predictions. For example, Fig. \ref{fig: correlations} shows how  $\textrm{BR}(\mu\rightarrow 3e)$ and $\textrm{CR}(\mu\rightarrow e, \textrm{Al})$ are correlated differently for two  representative FN Dirac textures.  These textures were chosen because they predict very similar \emph{average} values for both observables, but their widely diverging distributions makes it possible for two measurements to discriminate between these possibilities. Further correlation plots are included in the ancillary materials.

\noindent \textbf{Neutrino Mass Ordering --}
Most realistic natural textures in all scenarios favor normal ordering (NO). Inverted ordering (IO) is preferred by only a small fraction of Majorana textures and a somewhat larger subset of seesaw textures. Within the FN framework, an experimental determination of IO  (e.g. by 
DESI measuring $\sum m_\nu$ with a precision of 0.02 eV~\cite{DESI:2016fyo})
would therefore strongly disfavor Dirac neutrinos, while still allowing for Majorana or type-I seesaw scenarios.

\noindent \textbf{$\bm{0\nu\beta\beta}$ Decay --}
Predictions for $0\nu\beta\beta$ for 100 of the most realistic natural textures in Majorana and type-I seesaw models are shown in Figure~\ref{fig: seesaw 0nuBB}. For both scenarios, textures with inverted ordering (IO) consistently feature values of 
$m_{ee} = |\sum_i m_i V^2_{ei} |$
at the very lower edge of the currently allowed range.
We emphasize that this is not imposed on our scan and constitutes a genuine prediction of the FN mechanism -- while all shown analyses impose the cosmological bound on $\sum m_\nu$~\cite{ParticleDataGroup:2022pth}, there are no meaningful changes if we instead impose the larger laboratory bound~\cite{Katrin:2024tvg}.
This IO prediction lies well within the capabilities of next-generation $0\nu\beta\beta$ experiments. 
For textures yielding normal ordering (NO) there are significant phenomenological differences between FN  Majorana and type-I seesaw. The latter remains unlikely to be detected in upcoming laboratory searches -- barring a few exceptional textures -- while  normal-ordered FN Majorana scenarios predict $m_{ee}$ to lie well within or at most one order of magnitude below the sensitivity of upcoming laboratory searches. This opens up the tantalizing prospect of either detecting $0\nu\beta\beta$ or strongly disfavoring the entire Majorana  FN framework.

\section{Conclusions and Outlook}
\label{sec:conclusions}
In this work, we have systematically explored FN models in the leptonic sector, identifying realistic natural textures for Dirac, Majorana and type-I seesaw neutrino mass generation mechanisms. Our study also sheds light on related scenarios, including type-II seesaw cases (see Appendix \ref{app: uv completion majorana}). By extending the FN framework to leptons, we demonstrated that Dirac and Majorana FN models predict distinct correlations in detectable CLFV rates, providing characteristic signatures that set them apart from anarchic scenarios. Experimental signals are most likely to appear in the muon sector, with $\mu \rightarrow e \gamma$, $\mu \rightarrow 3e$, and $\mu$-$e$ conversion on nuclei as the most promising channels for future probes. Our results also motivate new conversion experiments in a variety of targets, most notably gold in addition to the already planned aluminum~\cite{Mu2e:2014fns}.
FN models with Majorana and type-I seesaw neutrinos offer testable predictions for $0\nu\beta\beta$ decay experiments, and a determination of inverted neutrino mass ordering would disfavor Dirac FN models.
Our conclusions are unaffected by adopting either cosmological or laboratory bounds on $\sum m_\nu$.

Our analysis constitutes the most model-exhaustive study of the Froggatt-Nielsen mechanism in the lepton sector to date, demonstrating both their universal predictive power and their capacity to diagnose the physics underlying lepton flavor.  
We anticipate that detection prospects will be greatly enhanced by considering observables involving flavor violation in both quarks and leptons, which motivates a joint analysis of the FN mechanism in both the quark and lepton sectors simultaneously. Integrating FN models with extended scalar sectors, such as multi-Higgs models, may also yield novel phenomenological insights in our model-exhaustive approach. \\

\textbf{Note Added:} While this paper was being finalized, ref. \cite{Ibe:2024cvi} appeared which takes a more literally Bayesian method to identify realistic and natural FN charge assignments for quark and lepton sectors. This analysis adopts a broadly similar approach to our work, including considering both Majorana and type-I seesaw models. Their work includes larger maximum flavor charges and considers FN in both the quark and lepton sector, which allows them to consider nucleon decay observables. On the other hand, our lepton sector FN analysis is more general in the range of allowed $\epsilon$ values, nonzero flavor charges for RH neutrinos, the inclusion of the Dirac neutrino case, and our predictions for CLFV observables. Our analyses are therefore highly complementary.

\section*{Acknowledgments}
The work of CC was partially supported by the Cluster of Excellence \textit{Precision Physics, Fundamental Interactions, and Structure of Matter} (PRISMA$^+$, EXC 2118/1) within the German Excellence Strategy (Project-ID 390831469). CC would also like to thank Perimeter Institute for hospitality during the completion of this work.  This research was supported in part by Perimeter Institute for Theoretical Physics. Research at Perimeter Institute is supported by the Government of Canada through the Department of Innovation, Science and Economic Development and by the Province of Ontario through the Ministry of Research, Innovation and Science.
The work of DC and MM was supported in part by Discovery Grants from the Natural Sciences and Engineering Research Council of Canada (NSERC), the Canada Research Chair program, the Alfred P. Sloan Foundation, the Ontario Early Researcher Award, and the University of Toronto McLean Award.
The work of MM was also supported in part by the NSERC Postgraduate Scholarship-Doctoral.
Fermilab is operated by the Fermi Research Alliance, LLC under Contract DE-AC02-07CH11359 with the U.S. Department of Energy. This material is based partly on support from the Kavli Institute for Cosmological Physics at the University of Chicago through an endowment from the Kavli Foundation and its founder Fred Kavli. 

\appendix


\section*{ --- APPENDIX --- }

\section{Details of Numerical Analysis}\label{app: numerics}

\subsection{Identifying Realistic Natural FN Textures }\label{app: detailed methods SM fit}

This section provides details on the method used to identify  natural realistic FN textures for leptons.

We allow for FN charges up to 7 for charged leptons, 7 for RH neutrinos in type-I seesaw, and 9 for RH neutrinos in the Dirac case. These large charge values are necessary to produce the observed neutrino masses for $\epsilon \sim 0.1$. 

Permutations of fields within a family do not represent physically distinct models. To remove these redundancies, we impose the following convention for the FN charges $X_\alpha$, $\alpha \in \{L, e, N \}$: $|X_{\alpha_i}| \geq |X_{\alpha_j}| $ for $i <j$ when all $X_\alpha$ have the same sign, otherwise $X_{\alpha_i} \geq X_{\alpha_j}$. We also remove mirror charges, which are related by multiplying all charges by $-1$ and reordering them.  

Each texture generates masses and mixings, which are compared to experimental values. Specifically, the observables we consider are the charged lepton masses, the PMNS matrix elements $|V_{11}|$, $|V_{13}|$, $|V_{23}|$, $|V_{21}|$, $|V_{31}|$, and $|V_{32}|$\footnote{Six PMNS elements are included because the uncertainties on these parameters are large enough that fitting only three independent elements, as would typically suffice for a unitary matrix, can still produce mixing matrices that deviate significantly from the measured values of the remaining elements.},  the neutrino mass squared differences, $\Delta m_{21}^2=m_2^2 - m_1^2$ and $\Delta m_{32}^2=m_3^2 - m_2^2$, and the cosmological bound on the sum of neutrino masses. To account for scenarios where this bound may be relaxed~\cite{Oldengott:2019lke,Chacko:2019nej,Escudero:2022gez}, we also repeated our analyses using the laboratory bound~\cite{Katrin:2024tvg}. The results show no significant differences. 
The measured values and uncertainties for these parameters are listed in Table \ref{tab: measured parameters}. For each observable $\mathcal{O}$, the fractional deviation from the experimental value is defined as 
\begin{equation}\label{eq:xi_exp_O_appendix}
\delta_\mathcal{O} = 
\begin{cases} 
\exp \left|\ln{\left(\frac{\mathcal{O}^{\rm FN}}{ \mathcal{O}^{\rm exp}_\mathrm{min}} \right)} \right| & \text{if } \mathcal{O}^{\rm FN} < \mathcal{O}^{\rm exp}_\mathrm{min}, \\[6pt]
1 & \text{if } \mathcal{O}^{\rm exp}_\mathrm{min} \leq \mathcal{O}^{\rm FN} \leq \mathcal{O}^{\rm exp}_\mathrm{max}, \\[6pt]
\exp \left|\ln{\left(\frac{\mathcal{O}^{\rm FN}}{\mathcal{O}^\mathrm{exp}_\mathrm{max}}\right)} \right| & \text{if } \mathcal{O}^{\rm FN} > \mathcal{O}^{\rm exp}_\mathrm{max}.
\end{cases}
\end{equation}
Here $\mathcal{O}^{\rm exp}_{\rm min}$ and $\mathcal{O}^{\rm exp}_{\rm max}$ are the experimental lower and upper bounds for the observable $\mathcal{O}$, respectively.
We then scan over many possible coefficient choices $c_{ij}^\alpha$ as explained in the main text. Textures can then be ranked by
$\mathcal{F}_x$ for different $x$. 

\begin{table}
    \centering
    \begin{tabular}{c|c|c}
       Parameter  &  Exp. Value & Ref.\\ \hline

        $m_e$(1 TeV)  & $0.489535765^{+0.000000013}_{-0.000000012}$ MeV & \cite{Xing:2011aa} \\
        $m_\mu$(1 TeV) & $103.3441945 \pm 0.0000059$ MeV & \cite{Xing:2011aa} \\
        $m_\tau$(1 TeV) & $1756.81 \pm 0.16$ MeV & \cite{Xing:2011aa} \\
       $\Delta m_{21}^2 $ & $(7.53 \pm 0.18 ) \times 10^{-5}\ \text{eV}^2$ & \cite{ParticleDataGroup:2022pth} \\
       $\Delta m_{32}^2 $ (IO) & $(-2.536 \pm 0.034) \times 10^{-3}\ \text{eV}^2$ & \cite{ParticleDataGroup:2022pth} \\
       $\Delta m_{32}^2 $ (NO) & $(2.453 \pm 0.033) \times 10^{-3}\ \text{eV}^2$ & \cite{ParticleDataGroup:2022pth} \\
       $\sum m_\nu$ (cosmo) & $\leq 0.12$ eV (95\% CL) & \cite{ParticleDataGroup:2022pth} \\
       $\sum m_\nu$ (KATRIN) & $\leq 1.35$ eV (90\% CL) & \cite{Katrin:2024tvg} \\
       $|V_{12}|$ & $[0.513, 0.579]$ & \cite{Esteban:2020cvm} \\
       $|V_{13}|$ & $[0.143,0.155]$ & \cite{Esteban:2020cvm} \\
       $|V_{23}|$ & $[0.637,0.776]$ & \cite{Esteban:2020cvm} \\
       $|V_{21}|$ & $[0.234,0.500]$ & \cite{Esteban:2020cvm} \\
       $|V_{31}|$ & $[0.271,0.525]$ & \cite{Esteban:2020cvm} \\
       $|V_{32}|$ & $[0.477,0.694]$ & \cite{Esteban:2020cvm} \\
    \end{tabular}
    \caption{Measured parameters and their uncertainties. Charged lepton masses are evaluated at a scale of 1 TeV and PMNS elements are given in $3\sigma$ ranges. We conduct all analyses twice, either imposing the cosmology or laboratory bound on $\sum m_\nu$, and find no meaningful difference in our conclusions.}
    \label{tab: measured parameters}
\end{table}

\subsection{Increasingly Focused Scan Sequence}

For each mass mechanism, the numerical procedure described above is performed in several stages. This is necessary because the full parameter space is enormous: for example, in the Dirac case there are over 100 million distinct charge assignments $\{X_L,X_e,X_N \}$. 

Scanning all textures in full (i.e. with a large number of trials) is numerically prohibitive, so we implement  preliminary scans to discard unpromising textures early. A distinct numerical advantage of our approach is that, generically, relatively large values of $\mathcal{F}_x$ for different choices of $x$ are highly correlated for natural realistic textures. This correlation allows us to perform an initial scan with few trials per texture, applying a cutoff at $\mathcal{F}_5$. Textures surpassing this initial threshold are then subjected to more intensive scrutiny for many more trials, focusing on $\mathcal{F}_x$ values for $x$ very close to 1. This strategy not only conserves computational resources but also ensures that only the most promising textures are examined in depth. We validate this method a posteriori by confirming that it consistently identifies textures capable of providing exact fits to the data, as detailed in Appendix~\ref{app: natural_fits}.

 \noindent \textbf{Charged lepton prescan --} 
 The first stage targets the charged lepton sector, where we start by examining 231,232 distinct charge assignments $\{X_L,X_{\bar e}\}$. This number can be reduced to 110,590 unique charge difference matrices $n_{ij}^\ell$, as permutations and redundancies are removed. Matrices containing a zero difference produce an eigenvalue of the same order as the Higgs VEV, which is too large for the tau mass. Removing these yields $\sim$66k unique matrices, corresponding to $\sim$134k textures.
We perform a charged lepton-only scan by running 1000 coefficient trials per texture and computing 
$\mathcal{F}_x$ using only the charged lepton masses as observables.
Textures are retained if they satisfy $\mathcal{F}_5>0$, meaning that at least a small fraction of trials match the charged lepton data.
At this stage, $\sim$ 46k charge assignments $\{X_L,X_{\bar e}\}$ pass the  $\mathcal{F}_5>0$ cut.

\noindent \textbf{Secondary preselection} -- Next, we incorporate constraints from the neutrino sector and impose additional cuts, which depend on the specific mass mechanism. 
\begin{itemize}[itemsep=0pt]
 \item {\bf Dirac --}  For the Dirac case, since neutrino masses depend only on $\epsilon$, the required charge differences $n^\nu$ must produce the appropriate level of suppression, $\epsilon^{n^\nu} \sim m_\nu$. For each charged lepton texture $\{X_L,X_{\bar e}\}$, we determine the allowed RH neutrino charges by identifying the minimal viable value of $\epsilon$ from the charged lepton prescan, denoted $\epsilon_{\rm min}$, that meets the criteria $\Xiexp<5$. We then use $\epsilon_{\rm min}$ to compute the minimum charge difference ${n^\nu}$ that satisfies a conservative bound on the total neutrino mass (e.g. $m^\nu_{\rm lim} = 100 \sum m_\nu$).
    Any texture with charge differences ${n^\nu}$ below this minimum is discarded.  
    This procedure drastically reduces the number of candidate textures $\{X_L,X_{\bar e},X_N\}$ to $\sim$ 74k, corresponding to $\sim$ 35k unique charge difference matrices $\{n^\ell,n^\nu\}$.
    \item {\bf Majorana --} In the Majorana scenario, there are no RH neutrino charges, so the neutrino charge differences $n^W$ are determined directly from the LH charges $X_L$.  After the charged lepton prescan, the top-performing textures are directly passed to the full scan without further cuts. The number of candidate charge difference matrices $\{n^\ell, n^W \}$ for the Majorana case is $\sim$94k. 
    \item {\bf Seesaw --} For type-I seesaw, the mass and mixing parameters depend on three charge difference matrices: $n^\ell$ (charged lepton Dirac), $n^\nu$ (neutrino Dirac), $n^N$ (RH Majorana). The dependence of the neutrino masses on $\Lambda$ prevents us from using cuts on $\epsilon_{\rm min}$ as was done in the Dirac case. 
    Due to the enormous number of possibilities (over $64$ million charge assignments for maximum charge 7, after the charged lepton prescan), a preliminary scan is performed. Specifically, we run a very coarse scan with 40 trials per texture, requiring $\mathcal{F}_5>10\%$. Only textures passing this criterion are subjected to the full scan with 1000 trials. 
    This procedure retains $\sim$ 2.7 million textures for further analysis. Despite the looser cuts, we expect this approach to capture all relevant textures with $\mathcal{F}_5>50\%$. 
\end{itemize}

\noindent \textbf{Final scans} -- 
After the secondary preselection, we perform two more scans for all mass mechanisms. First, for each remaining texture, we do a fresh scan with 1000 coefficient choices and rank textures based on $\mathcal{F}_2$. Textures with the same $\mathcal{F}_2$ are ranked by $\mathcal{F}_5$. 
Finally, for the top  1000 textures from this prescan we do a final scan with $10^5$ random coefficient choices. Textures are ranked by $\mathcal{F}_x$ with $x \approx 1.2 - 1.3$ depending on the FN scenario. The exact value of $x$ is chosen to be as close to 1 as possible while optimizing the statistical significance of the ranking.

Our phenomenological analysis focuses solely on observables related to lepton flavor or lepton number violation. Specifically we include LFV processes at low and high energies and neutrinoless double $\beta$ decay.  
The complete list of observables, alongside their current and projected experimental sensitivities, are summarized in Table \ref{tab: experimental measurements}. 

\begin{table}[t]
    \centering
    \begin{tabular}{c|c|c|c|c}
        Observable & Current & Ref. & Future & Ref. \\ \hline
BR$(\mu^+\rightarrow e^+ \gamma)$ & $ 4.2 \times 10^{-13}$ & \cite{MEG:2016leq} & $6\times 10^{-14}$ & \cite{Baldini:2013ke}\\
BR$(\mu^+\rightarrow e^+ e^- e^+)$ & $ 1.0 \times 10^{-12}$ & \cite{SINDRUM:1987nra} & $10^{-16}$ & \cite{Blondel:2013ia}\\
BR$(\tau\rightarrow e \gamma)$ & $ 3.3 \times 10^{-8}$ & \cite{BaBar:2009hkt} & $2 \times 10^{-9}$ & \cite{Belle-II:2018jsg}\\
BR$(\tau\rightarrow \mu \gamma)$ & $ 4.2 \times 10^{-8}$ & \cite{Belle:2021ysv} & $10^{-9}$ &  \cite{Belle-II:2018jsg}\\
BR$(\tau\rightarrow eee)$ & $2.7 \times 10^{-8}$ & \cite{Hayasaka:2010np} & $4\times 10^{-10}$ & \cite{Belle-II:2018jsg}\\
BR$(\tau\rightarrow \mu\mu\mu)$ & $2.1 \times 10^{-8}$ & \cite{Hayasaka:2010np}& $4\times 10^{-10}$ &\cite{Belle-II:2018jsg} \\
 BR$(\tau^-\rightarrow \mu^+e^-\mu^-)$ & $2.7 \times 10^{-8}$ & \cite{Hayasaka:2010np}& $4\times 10^{-10}$ &\cite{Belle-II:2018jsg} \\
BR$(\tau^-\rightarrow e^+\mu^-\mu^-)$ & $1.7 \times 10^{-8}$ & \cite{Hayasaka:2010np} & $3\times 10^{-10}$ &\cite{Belle-II:2018jsg}\\
BR$(\tau^-\rightarrow e^+\mu^-e^-)$ & $1.8 \times 10^{-8}$ & \cite{Hayasaka:2010np} & $3\times 10^{-10}$ &\cite{Belle-II:2018jsg} \\
BR$(\tau^-\rightarrow \mu^+e^-e^-)$ & $1.5 \times 10^{-8}$ & \cite{Hayasaka:2010np} & $3\times 10^{-10}$ &\cite{Belle-II:2018jsg}\\
 CR$(\mu^- \text{Ti}\rightarrow e^- \text{Ti})$ & $6.1\times 10^{-13}$ & \cite{Wintz:1998rp}&-&- \\
CR$(\mu^- \text{Pb}\rightarrow e^- \text{Pb})$ & $4.6\times 10^{-11}$ & \cite{SINDRUMII:1996fti}&-&-\\
CR$(\mu^- \text{Au}\rightarrow e^- \text{Au})$ & $7.0\times 10^{-13}$ & \cite{SINDRUMII:2006dvw}&-&-\\
CR$(\mu^- \text{Al}\rightarrow e^- \text{Al})$ & - & - & $3\times 10^{-17}$ &  \cite{Mu2e:2014fns}\\
$m_{ee}$ & 36 meV & \cite{ParticleDataGroup:2022pth} & 3 meV & \cite{nEXO:2021ujk,CUPID:2022wpt} \\
$\mu\mu \rightarrow e \mu$ & - & - & - & -\\
$\mu\mu \rightarrow \mu\tau$ & - & - & - & -\\
$\mu\mu \rightarrow e \tau$ & - & - & - & -\\
\end{tabular}
\caption{Current experimental limits and future sensitivities for the observables considered in our analysis. Current limits are given at $90\%$ confidence level, except for the bound on 
$m_{ee}$, which is marginalized over different nuclear matrix element values with \Oone\, uncertainty.}
    \label{tab: experimental measurements}
\end{table}

\begin{table}[t]
\renewcommand{\arraystretch}{1.3} 
\centering
\begin{tabular}{c|c|c|c}
\hline 
$Q_{LL}$  & $\left( \bar L_p \gamma_\mu L_r \right)\left(\bar L_s \gamma^\mu L_t \right)$ & $Q_{LQ}^{(1)}$ & $\left( \bar L_p \gamma_\mu L_r \right)\left(\bar Q_s \gamma^\mu Q_t \right)$ 
\\
$Q_{ee}$  & $\left( \bar e_p \gamma_\mu e_r \right)\left(\bar e_s \gamma^\mu e_t \right)$  & $Q_{LQ}^{(3)}$ &  $\left( \bar L_p \gamma_\mu \tau^I L_r \right)\left(\bar Q_s \gamma^\mu \tau^I Q_t \right)$ 
\\
 $Q_{Le}$  & $\left( \bar L_p \gamma_\mu L_r  \right)\left(\bar e_s \gamma^\mu e_t \right)$ &   $Q_{He}$ &  $\left(H^\dagger i \overleftrightarrow D_\mu H\right) \left( \bar e_p \gamma_\mu e_r \right)$  
 \\
$Q_{ed}$ & $\left( \bar e_p \gamma_\mu e_r \right)\left(\bar d_s \gamma^\mu d_t \right)$ &  $Q_{HL}^{(1)}$ & $\left(H^\dagger i \overleftrightarrow D_\mu H\right) \left( \bar L_p \gamma_\mu L_r \right)$
\\
$Q_{eu}$ & $\left( \bar e_p \gamma_\mu e_r \right)\left(\bar u_s \gamma^\mu u_t \right)$ & $Q_{HL}^{(3)}$ & $\left(H^\dagger i \overleftrightarrow D^I_\mu H\right) \left( \bar L_p \tau^I \gamma_\mu L_r \right)$
\\
$Q_{Lu}$ & $\left( \bar L_p \gamma_\mu L_r \right)\left(\bar u_s \gamma^\mu u_t \right)$ & $Q_{eW}$ & $\left( \bar L_p \sigma^{\mu\nu} e_r \right)\tau^I H W^I_{\mu\nu}$
\\
$Q_{Ld}$ & $\left( \bar L_p \gamma_\mu L_r \right)\left(\bar d_s \gamma^\mu d_t \right)$  &  $Q_{eB}$ & $\left( \bar L_p \sigma^{\mu\nu} e_r \right)H B_{\mu\nu}$
\\
$Q_{Qe}$ & $\left( \bar Q_p \gamma_\mu Q_r \right)\left(\bar e_s \gamma^\mu e_t \right)$  &  

\end{tabular}
\caption{Warsaw-basis SMEFT operators included in our analysis \cite{Grzadkowski:2010es}. 
$SU(2)_L$ indices are indicated by $j,k=1,2$ where necessary.
Flavor indices are given by $p,r,s,t$=$1,2,3$. For leptons we include all flavor combinations. For quarks we only include flavor-diagonal combinations of the three lightest quarks, as the corresponding Wilson coefficients are enough to capture the leading contribution to $\mu-e$ conversion in nuclei. All other operators are set to zero.}
\label{tab: SMEFT operators}
\end{table}
\subsection{Predicting Lepton Violation in FN}\label{app: detailed methods pheno}

For the predictions, we employ the SMEFT framework. We construct effective operators following the power counting dictated by formal \UFN invariance, as in

\beq
    \mathcal{O}_4 = \frac{c_{ijkl}}{\Lambda^2}  \left(\Bar{\psi}_i \psi_j \right) \left(\Bar{\psi}_k \psi_l \right) \epsilon^{n_{ijkl}},
\eeq
where $c_{ijkl}$ are ${\cal O}(1)$ coefficients and we have defined
\be
n_{ijkl} \equiv |X_{\psi_i} - X_{\psi_j} +X_{\psi_k} -X_{\psi_l}|.
\ee

We assume all Wilson coefficients to be \Oone\,, and generate them randomly with the same priors used for the effective Yukawa coefficients. For simplicity, we consider only SMEFT operators that contribute at tree-level to our observables (we checked explicitly with {\tt wilson}~\cite{Aebischer:2018bkb} that running effects do not affect our results significantly). We collect these operators in Table~\ref{tab: SMEFT operators}.

To calculate low-energy LFV processes, such as muon and tau decays and muon-to-electron conversion in nuclei, we use {\tt flavio} \cite{Straub:2018kue}.
For high-energy collider observables, we follow Ref.~\cite{Paraskevas:2018mks}. As for $0\nu\beta\beta$, the effective Majorana mass $m_{ee}$ is given by $ m_{ee} = \left|\sum_i m_i V^2_{ei} \right|$.

To generate predictions for the top FN textures, we  retain all trials of  our final scan that meet the goodness of fit criterium $\Xiexp\leq 2$, storing the associated rotation matrices, $\epsilon$, and, where applicable, $\Lambda_{(W)}$.\footnote{We explicitly checked with the top few textures that performing exact fits, as detailed in Appendix \ref{app: natural_fits}, and using these exact fits as input for the phenomenological study did not significantly alter the results.}  Trials are then ranked by $\Xiexp$, and for the top 500, we calculate the suppression for the Wilson coefficients of the operators in Table \ref{tab: SMEFT operators}. Note that to remain as model-independent as possible we have excluded  operators where the two quark fields have different chiralities, as these could be further  suppressed if, for instance, the quarks carried FN charges. However, we explicitly verified that including these operators results in only a $\sim 5\%$ correction. We only included flavor diagonal quark operators for the same reason. After rotating to the mass basis, we derive the predictions for the observables in \ref{tab: experimental measurements}. 
For each observable we find the average and standard deviation and compare it to data. This allows us to establish a lower bound on $\Lambda$ for Dirac and Majorana scenarios, and to check whether the predictions for type-I seesaw models comply with current bounds. Note that our bounds on $\Lambda$ are derived purely from processes involving SM states. Considering the phenomenology of the flavon itself may yield additional, but model-dependent, bounds \cite{Heikinheimo:2018luc}.

\subsection{Natural Fits}\label{app: natural_fits}

\begin{figure*}[t]
\centering
\includegraphics[scale=0.33]{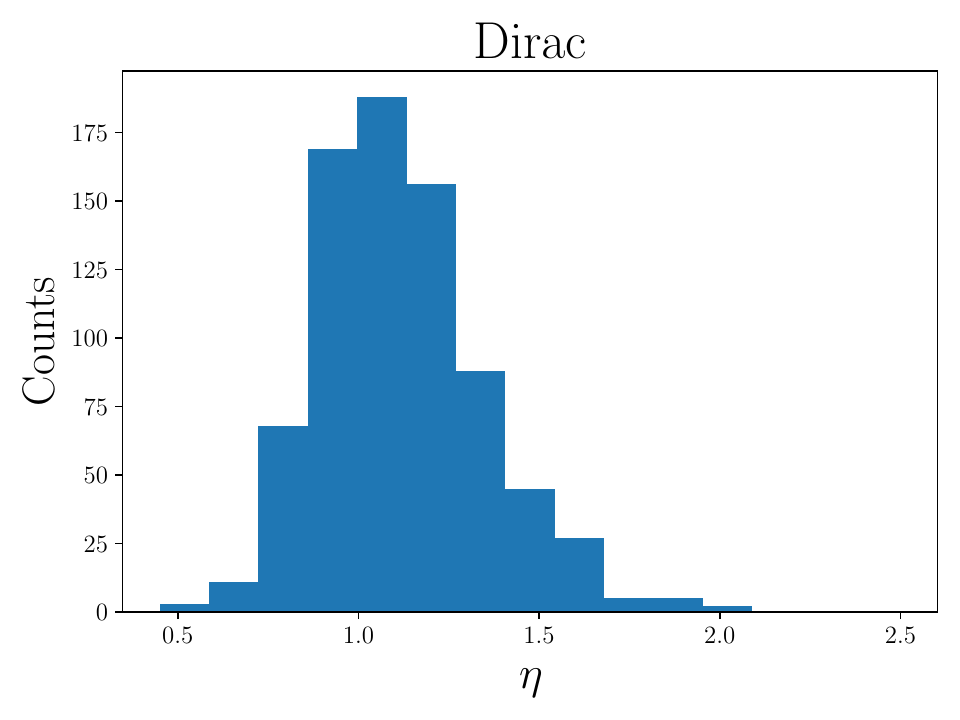}
\includegraphics[scale=0.33]{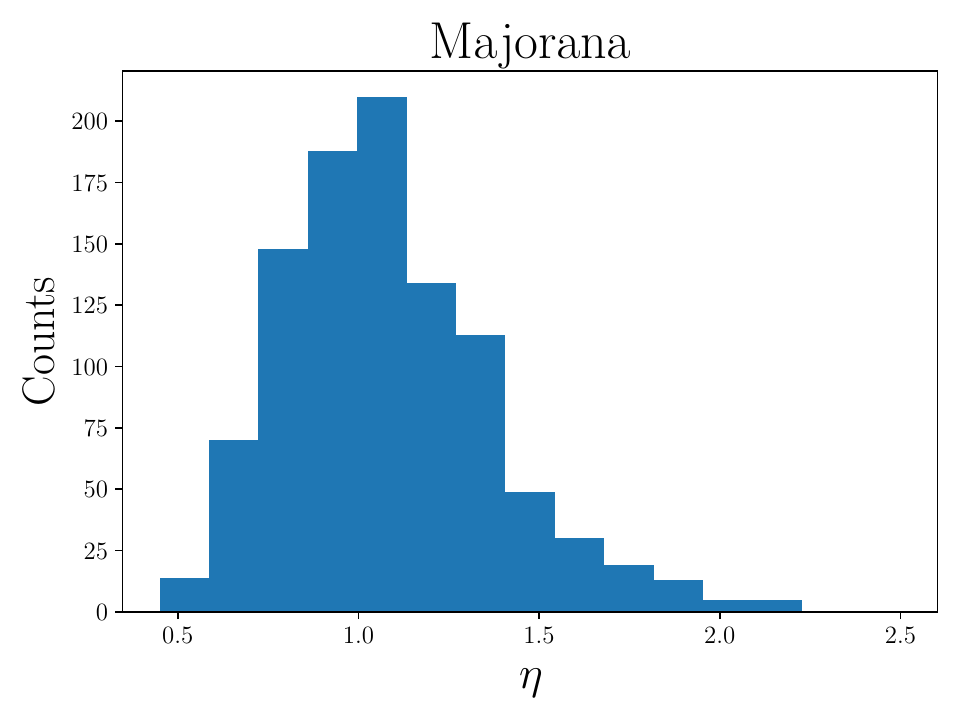}
\includegraphics[scale=0.33]{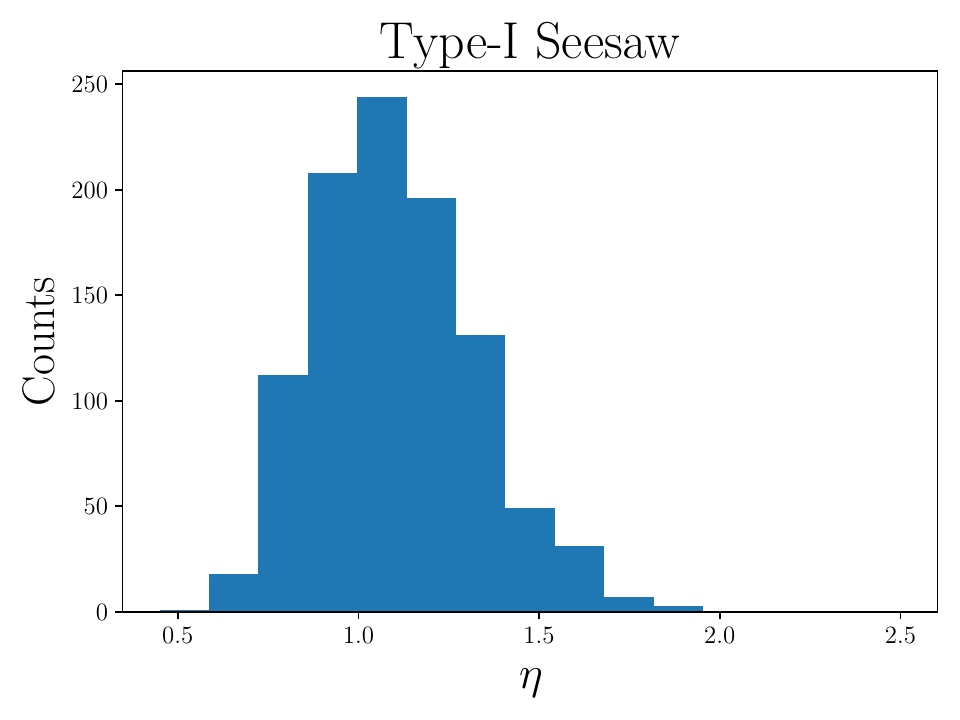}
\caption{Histograms of the $\eta$ parameter distribution for $1000$ trials that yield exact fits to the measured leptonic masses and mixings parameters. For each neutrino mass mechanism we show the result for the top-performing textures: for Dirac neutrinos (left), $X_L = \{6,5,5 \}$, $X_e=\{-3,-2,0\}$, $X_N = \{9,8,8 \}$; for Majorana (middle), $X_L = \{2,0,-1 \}$, $X_e=\{7,6,4\}$; and for type-I seesaw (bottom), $X_L = \{6,1,-1\}$, $X_e=\{7,7,6\}$, $X_N = \{3,0,-4 \}$. 
}
\label{fig: sm fits}
\end{figure*}

This section details the method for adjusting the coefficients $c_{ij}^\alpha$ ($\alpha = \ell, \nu, W, N$) to reproduce exactly the masses and mixing parameters in Table \ref{tab: measured parameters}. Our goal is to demonstrate that with minimal adjustments to the initial \Oone\ coefficients, the predictions for our top FN textures can be brought into full agreement with data. 

For each mass generation mechanism, we focus on the top texture as ranked in the main text. We use a simplex minimization algorithm
to refine the coefficients $c^\alpha_{ij}$ to fit the experimental values by minimizing the cost function in \Eq{eq:xi_exp_O_appendix}. We do this using as starting point 1000 coefficient sets from trials that yielded a promising $\Xiexp<2$. 

To assess the extent of the adjustment needed, we introduce the parameter:
\beq
    \eta = \log_{10} \frac{\max \left| c \right|}{\min \left| c \right|}
\eeq
where $c$ spans all the coefficients in the effective Yukawa matrices $c^\ell,\ c^\nu,\ c^W$, and $c^N$. 
Histograms of $\eta$ (Fig. \ref{fig: sm fits}) indicate that typically only a modest range variation is required, suggesting our selected textures reproduce experimental observations naturally, without extensive tuning or significant adjustments of coefficients. The examination of individual coefficient distributions further confirms that they adhere closely to our \Oone\, assumption.
Overall, this check confirms our method of identifying natural and realistic textures without doing explicit fits to the experimental data.

\section{More on Phenomenology and Supplemental Plots}\label{app: supplemental plots}

\begin{figure}[t]
    \centering
    \includegraphics[width=0.9\linewidth]{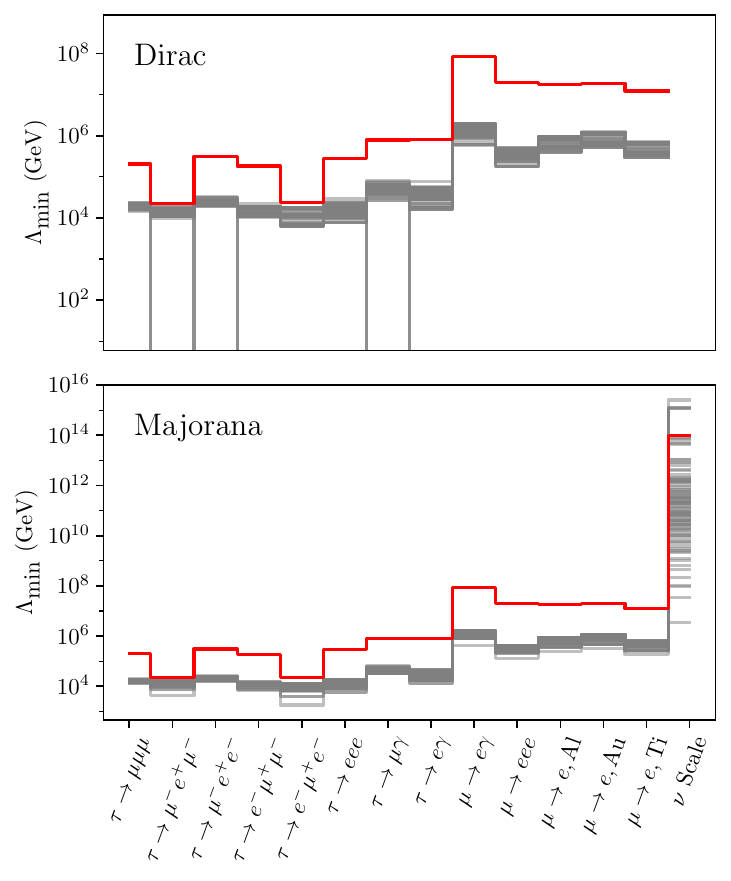}\\
    \caption{Bounds on the FN scale $\Lambda$ from current constraints on the processes shown on the horizontal axis, for the Dirac (top) and Majorana (bottom) FN scenarios. For the latter we also show the constraint from active neutrino masses if $\Lambda = \Lambda_W$.  }
    \label{fig:lowE_bounds}
\end{figure}

\begin{figure}[t]
    \centering
\includegraphics[width=0.9\linewidth]{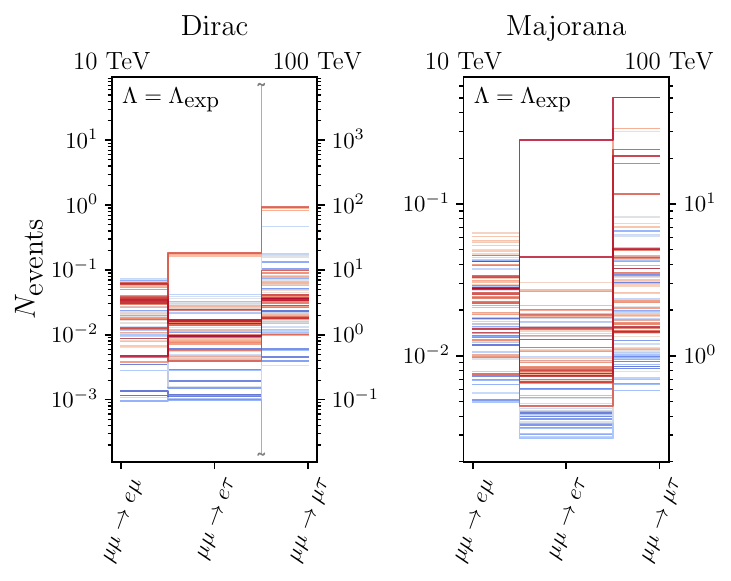} 
    \caption{ 
    Predicted muon collider event rates for CLFV processes in Dirac FN and Majorana FN scenarios with $\Lambda = \Lambda_{\rm exp} \neq \Lambda_W$ chosen to saturate current experimental bounds at $\Lambda \sim 10^6$~GeV. assuming an integrated luminosity of 10 ab$^{-1}$.
    In this figure, the ranking of each texture is indicated by color, with red textures ranked higher than blue textures. 
     The corresponding results for the Majorana scenario with $\Lambda = \Lambda_W$ or type-I seesaw are not shown, as the predicted rates are many orders of magnitude smaller.}
    \label{fig:muon_bounds}
\end{figure}

This appendix completes the discussion in the main text, supplementing the results with additional figures. \\

\noindent \textbf{Low-energy CLFV and bounds on $\Lambda$ --}  For both the Dirac and Majorana cases, the bounds on the FN scale $\Lambda$ are set by current low-energy CLFV bounds. Fig.~\ref{fig:lowE_bounds} illustrates that the strongest bounds are mostly set by $\mu \to e \gamma$ (and secondly by $\mu \to 3e$), with the strongest bound on $\Lambda$ being remarkably consistent across different textures.  The difference between the null and FN textures arises because FN observables sensitive to left-right couplings are suppressed with respect to the anarchic case. 

\textbf{Future muon colliders --} Collider experiments provide a potential platform to test LFV processes predicted by FN models, especially at high energies where muon colliders~\cite{Accettura:2023ked} outshine $e^+e^-$ colliders~\cite{Shiltsev:2019rfl} due to their higher accessible energies and the resulting enhancement of cross-sections. Our analysis indicates that the Dirac and Majorana scenarios (with $\Lambda = \Lambda_{\text{exp}}$), show the most promising but still very modest potential for detectable signals. Detailed results for these scenarios are illustrated in Fig. \ref{fig:muon_bounds}.

In Dirac FN models, certain textures exhibit an enhancement in the $\mu\tau$ final state.  This enhancement is mainly due to the equality of $X_{\mu_L}$ and $X_{\tau_L}$ charges, which leads to significant contributions from unsuppressed LH four-fermion operators. As a result, $\mu\tau$ processes are more likely to be detected compared to $e\tau$ processes, which are generally suppressed by larger charge differences that diminish mixing.
As shown in Fig. \ref{fig:muon_bounds}, we find $\mathcal{O}(1)$ and $\mathcal{O}(100)$ events for a c.o.m. energy of 10 and 100 TeV, respectively.  
 
For Majorana neutrinos (with $\Lambda = \Lambda_\mathrm{exp}$), the predictions at colliders vary notably from those in the Dirac case. Here, the $e\mu$ and $\mu\tau$ final states are expected to have nearly similar magnitudes, reflecting the specific FN charge differences among LH fields. This similarity suggests that collider experiments could provide insights into the LH charge assignments in FN models. Still, collider prospects remain very limited, with only
a few textures giving $\mathcal{O}(1)$ events at an extremely hypothetical 100 TeV machine.

\begin{figure}[t]
    \centering
    \includegraphics[width=0.9\linewidth]{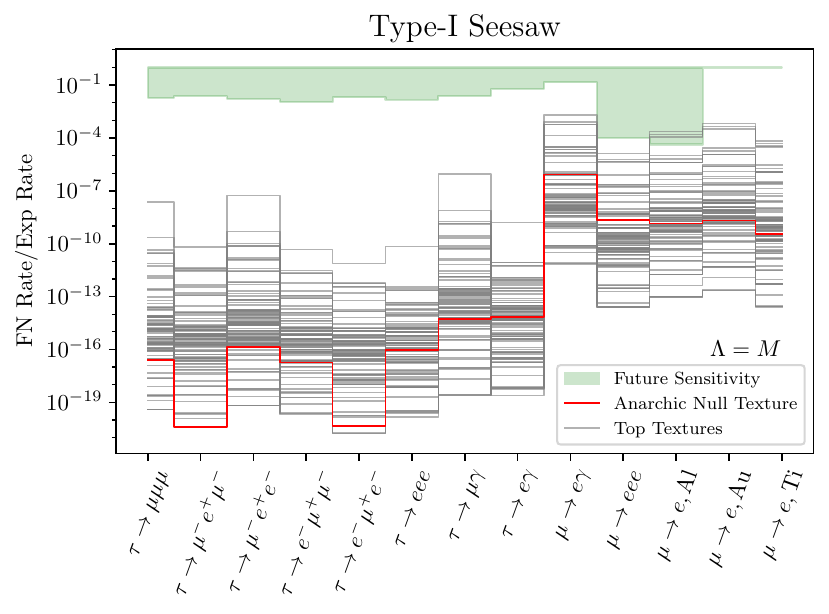}
    \caption{Average predicted CLFV decay rates for the 100 most realistic natural type-I seesaw FN textures (gray lines), relative to each observable's current constraint. 
    The FN scale $\Lambda$ is fixed by neutrino masses. .  
    Green shading indicates the reach of proposed future low-energy CLFV experiments, and the flavor-anarchic null texture is shown as a red  line for comparison. }
    \label{fig:seesaw low E observables}
\end{figure}

\textbf{Seesaw phenomenology -- }
The phenomenology of the top 100 FN type-I seesaw models, detailed in Fig.~\ref{fig:seesaw low E observables}, aligns broadly with that in the Dirac and Majorana cases (displayed in the main text), with less variability compared to the null texture. Unfortunately, none of the identified textures have a low enough scale to be probed by any currently planned CLFV experiment.

\begin{figure}[t]
    \centering
    \begin{tabular}{c}
    \includegraphics[width=0.9\linewidth]{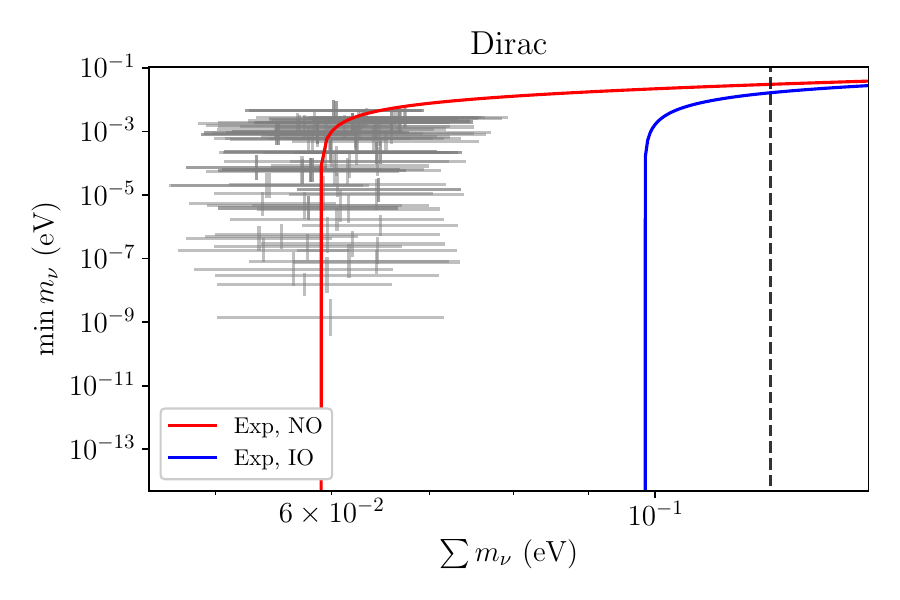} \\
    \includegraphics[width=0.9\linewidth]{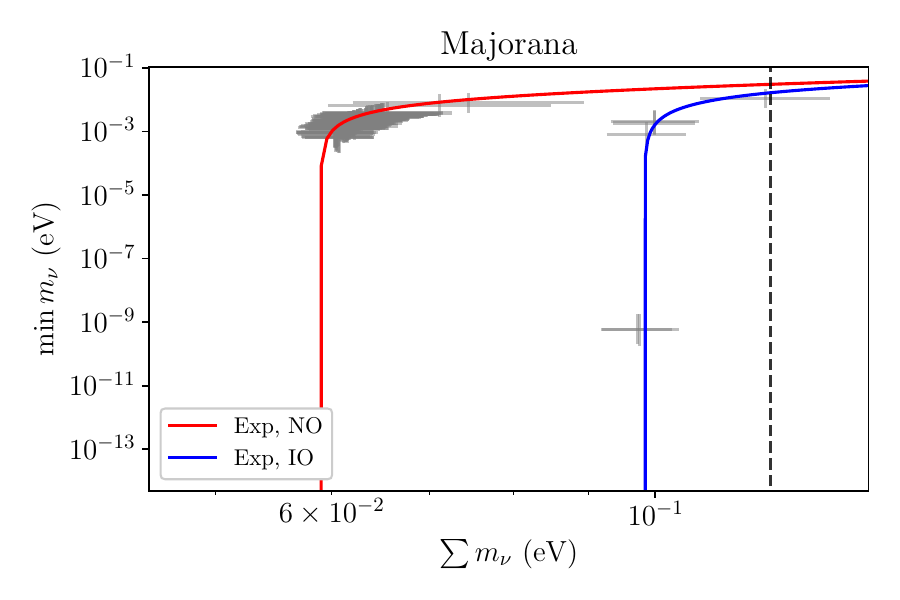}\\
    \includegraphics[width=0.9\linewidth]{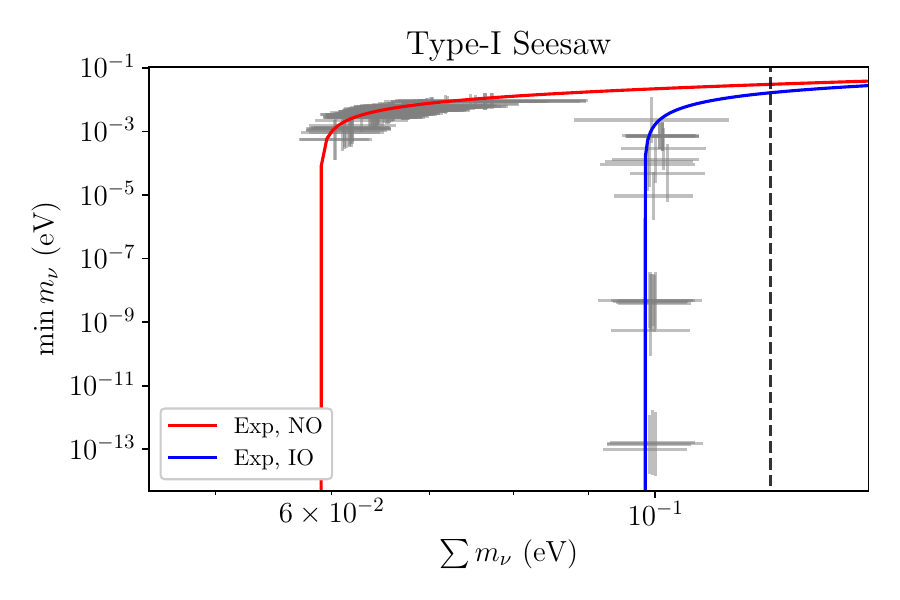} \\
    \end{tabular}
    \caption{Predictions for the lightest neutrino mass versus the sum of neutrino masses for the 100 most realistic textures in the Dirac (top), Majorana (middle) and type-I seesaw (bottom) FN scenarios. 
    The red (blue) line represents the mass of the lightest neutrino as implied by the values of $\Delta m_{21}^2$ and $\Delta m_{32}^2$ for a $\sum m_\nu$ in the NO (IO) case. The dashed line shows the current cosmological constraint on the sum of neutrino masses. Error bars show the $1\sigma$ spread of  predictions over 500 trials with $\Xiexp<2$. The large error bars on the Dirac scenario are due to the high powers of $\epsilon$ required to reproduce  neutrino masses, making this scenario more sensitive to small variations in $\epsilon$. }
    \label{fig: mlight msum}
\end{figure}

\textbf{Active Neutrino Masses} 
The predicted mass of the lightest neutrino and the sum of neutrino masses is shown in Fig. \ref{fig: mlight msum} for the top 100 textures of each mass mechanism. It is interesting to note that many of these natural realistic textures predict an ultralight active neutrino (see e.g.~\cite{Chacko:2016hvu, Alonso-Alvarez:2023bat}).

Dirac FN only predicts normal-ordered (NO) neutrinos. While DESI is projected to be able to measure the minimal NO scenario at $3\sigma$~\cite{DESI:2016fyo}, the resolution will not be sufficient to discriminate between Dirac textures. However, a positive determination of inverted-ordering (IO) would strongly disfavor the Dirac scenario.

\section{Generalizations}\label{app: generalizations}

\subsection{Non-Trivially Charged Higgs}\label{app: non trivially charged H}

One can also consider the case where the Higgs is non-trivially charged under \UFN. For the Dirac case, the Lagrangian is modified to:

\begin{multline}
    \mathcal{L} \supset -c^\ell_{ij}L_i H^\dagger \bar e_j  \epsilon^{|X_{L_i}+ X_{\bar e_j}-X_H|} \\
    -c^\nu_{ij}H L_i N_j \epsilon^{|X_{L_i}+X_{N_{j}}+X_H|}  + \text{h.c.}
\end{multline}

It is straightforward to identify textures with $X_H=0$ that give equivalent Yukawas. Requiring all fields of the same type to be shifted by the same amount (e.g. $X_{L_i}\rightarrow X_{L_i}+c$), there are three possible solutions, depending on  whether $X_L,\ X_e$ or $X_N$ is held fixed. Using $X$ to denote the original $X_H=0$ texture and $X^\prime $ to indicate the new $X_H\neq0$ texture, the three solutions are as follows:
 
\begin{equation}\label{eq: charged higgs dirac solutions}
    \begin{aligned}
        X_{L_i}^\prime &= X_{L_i}, ~~~ X^\prime_{e_j}= X_{\bar e_j}-X_H, ~~~X^\prime_{N_j}= X_{N_j}+X_H,\\
        X_{L_i}^\prime &= X_{L_i} + X_H, ~~~ X^\prime_{e_j}= X_{\bar e_j}-2 X_H, ~~~ X^\prime_{N_j}= X_{N_j}, ~~\\
        X_{L_i}^\prime &= X_{L_i} - X_H, ~~~ X^\prime_{e_j}= X_{\bar e_j}, ~~~ X^\prime_{N_j}= X_{N_j} +2 X_H.
    \end{aligned}
\end{equation}

For a given $X_H=0$ texture from our analysis, we can then easily find the corresponding $X_H\neq 0$ textures. In the Majorana and type-I seesaw cases, the requirement that the Weinberg operator or the Majorana RH neutrino mass term remains invariant selects the second solution in \Eq{eq: charged higgs dirac solutions}. 

The purely leptonic observables for the $X_H= 0$ texture and the equivalent $X_H\neq0$ textures remain the same. This can be seen by observing that most fermion bilinears are unaffected:

\begin{equation}
    \begin{aligned}
        (\bar \psi_i \psi_j)\epsilon^{|X^\prime_{\psi_j} - X^\prime_{\psi_i}|} &= (\bar \psi_i \psi_j)\epsilon^{|(X_{\psi_j} +c) - (X_{\psi_i}+c)|},\\
        &= (\bar \psi_i \psi_j)\epsilon^{|X_{\psi_j} - X_{\psi_i}|}.
    \end{aligned}
\end{equation}

Here, Dirac structures are omitted for clarity. As long as $\psi_i$, $\psi_j$ are of the same type (e.g. $\bar L L$ or $\bar e e$), the shift cancels in the exponent, and the bilinear remains invariant. Thus, all four-fermion operators in Table \ref{tab: SMEFT operators} are unchanged. Similarly, the operators $Q_{eW}$ and $Q_{eB}$ share the same form as the Yukawa couplings, and are therefore invariant by definition. The remaining operators contain bilinears of the form $H^\dagger H$, which are \UFN-neutral by construction. As all SMEFT operators we consider remain the same, the predictions for CLFV are unchanged.


\subsection{Other Sources of Lepton Number Violation}\label{app: lepton number violation}

In the type-I seesaw case, we have assumed no additional sources of lepton number violation beyond the Majorana mass of the RH neutrino. If this assumption is relaxed, the  Lagrangian is modified by an additional contribution to the Weinberg operator:

\begin{equation}
    \mathcal{L} \supset \mathcal{L}_{\rm SS} - \frac{c_{ij}^W \epsilon^{n^W_{ij}} }{\Lambda_{\cancel{L}}} (L_i H)(L_j H)\,, 
\end{equation}
where $\Lambda_{\cancel{L}}$ indicates the scale of the additional lepton number violation. 

The resulting neutrino masses can be derived similarly to those in the type-I seesaw discussed in the main text. We define the Weinberg mass as $m^2_{W,ij} = v^2 c_{ij}^W \epsilon^{n_{ij}^W}$ and note that the Weinberg operator does not contribute to the mixing between sterile and active eigenstates until $\mathcal{O}\left((v \Lambda^{-1})^3 \right)$~\cite{Dedes:2006ni}. The resulting neutrino masses are:

\beq\label{eq: type 1 seesaw nu mass w weinberg}
    m_\nu \approx  \frac{m^2_W}{\Lambda_{\cancel{L}}} - m_D M_N^{-1} m_D^T\,,
\eeq
where $m_D = v Y^\nu/\sqrt{2}$ and $M_N = M c^M \epsilon^{n^M}$. 

\begin{figure}[t]
    \centering
    \includegraphics[width=0.9\linewidth]{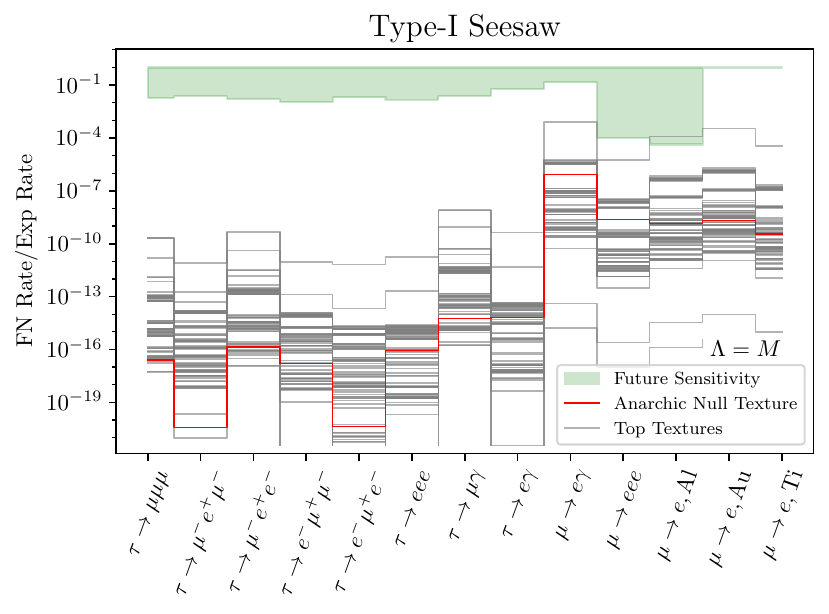}
    \caption{Phenomenology for the top 100 type-I seesaw textures if the FN sector is lepton number violating (with $\Lambda_{\cancel{L}}=\Lambda = M$), as in Fig. \ref{fig:seesaw low E observables}. The FN scale $\Lambda$ is fixed by neutrino masses. The axes are fixed to be the same as in Fig. \ref{fig:seesaw low E observables} for ease of comparison.  }
    \label{fig: seesaw w weinberg}
\end{figure}

We performed our analysis in the case where $\Lambda_{\cancel{L}}= \Lambda=M$, and found slightly different top textures. However, the overall conclusions remain unchanged: the scale of flavor violation for these models exceeds projected experimental sensitivities. The CLFV phenomenology for the top 100 textures is shown in Fig.~\ref{fig: seesaw w weinberg}.

\subsection{Type II Seesaw and other Majorana UV Completions}\label{app: uv completion majorana}

The Majorana case can arise from a variety of UV completions. For illustration, we consider the example of a FN-augmented type-II seesaw mechanism.
The type-II seesaw \cite{Magg:1980ut, Schechter:1980gr, Lazarides:1980nt, Mohapatra:1980yp, Wetterich:1981bx} has the form \cite{Ma:1998dx}

\be
    \mathcal{L}_{SS2} \supset  -\frac{y}{2} \Delta  {L} L -\mu_\Delta \Delta  
 H^\dagger 
H^\dagger 
    - M_\Delta^2 \text{Tr}[\Delta^\dagger \Delta],~~~
\ee 
where $\Delta$ is an electroweak scalar triplet with hypercharge $1$, and generation indices are omitted. The quantities $M_\Delta$ and $\mu_\Delta$ are dimensionful, and $y$ is dimensionless. When $M_\Delta$ is large, integrating out $\Delta$ yields a Weinberg operator with the prefactor:

\beq
    \frac{c^W_2}{\Lambda_W} = \frac{ 2 y \mu_\Delta }{M_\Delta^2}.
\eeq
Assuming that $\Delta$ is charged under $U(1)_{FN}$, the Lagrangian becomes 

\begin{multline}
        \mathcal{L}_{SS2} \supset -\frac{y_{ij}}{2}\epsilon^{|X_{Li}+X_{Lj}+X_\Delta|}L_i L_j \Delta \\
        - c_\mu \Lambda \epsilon^{|X_\Delta|} \Delta H^\dagger H^\dagger \\
        +\text{h.c.} 
        - (c_M\Lambda)^2 \text{Tr}[\Delta^\dagger \Delta].  
\end{multline}
where we have taken the mass scales $\mu_\Delta$ and $M_\Delta$ to be set by the flavon scale $\Lambda$, and $y,\ c_{\mu},\ c_{M}$ are \Oone\ prefactors. This implies the Weinberg coefficient

\beq
    \begin{aligned}
        \frac{c^W_2}{\Lambda_W} &\approx \frac{2 \left(y_{ij}\epsilon^{|X_{Li}+X_{Lj}+X_\Delta|} \right)\left( c_\mu \Lambda \epsilon^{|X_\Delta|}\right) }{(c_M\Lambda)^2} \\
        &\approx \frac{\tilde c_{ij}\epsilon^{n^W + 2 X_\Delta}}{\Lambda},
    \end{aligned}
\eeq
where $\tilde c \sim$ \Oone\,. In the second equality we have assumed all relevant charges to be positive, so that the absolute value can be dropped. As a result, the apparent scale of the Weinberg operator differs from the true FN scale by $\Lambda_W = \epsilon^{-2 X_\Delta} \Lambda$. Consequently, a given effective Weinberg scale can correspond to a much lower physical scale. However, if charges are such that the absolute values in the exponents cannot be dropped, different generations may have distinct FN suppression. This scenario cannot be mapped onto our generic Majorana textures and lies beyond the scope of this work. 

For the type-I seesaw, the same analysis yields: 

\begin{multline}
    \frac{c^W_2}{\Lambda_W}= \frac{1}{ \Lambda}  \left[ c^\nu_{ij}\epsilon^{|X_{L_i} + X_{N_{j}} |}\right.\\
    \left.\left(c^M_{jm}\epsilon^{|X_{N_{j}}+X_{N_{m}}|} \right)^{-1} 
    c^\nu_{mn}\epsilon^{|X_{L_n}+ X_{N_{Rm}}|} \right]\,. \nonumber
\end{multline}

As in type-II seesaw,  certain textures can be directly matched to a Majorana texture with a common scale. However, for some textures the effective scales are hierarchical and cannot be mapped onto the Majorana case as implemented in our method. Additionally, in some cases, a midscale UV completion of the Weinberg operator may generate additional contributions less suppressed than the FN-generated ones. In such cases, the phenomenological predictions would be somewhat different, depending on which operators are generated. These scenarios would require an ad-hoc study.

\section{Auxiliary Material}\label{app: aux material}

In the auxiliary material we include several additional documents for reference. The files
\begin{itemize}[itemsep=0pt]
    \item \texttt{dirac\_tex\_pheno\_cosmo.csv}
    \item \texttt{seesaw\_tex\_pheno\_cosmo.csv}   
    \item \texttt{majorana\_tex\_pheno\_cosmo.csv}
\end{itemize}
 provide phenomenological predictions for each neutrino mass generation model. 
Each file contains the charge assignments for the top 100 textures, along with the best values of $\epsilon$ and $\Lambda$, when relevant, averaged over trials with $\Xiexp<2$. 
NO indicates the fraction of trials that yield normal-ordered neutrinos. $\Lambda_\textrm{exp}$ is the scale implied by the most constraining observable. By default charge assignments are sorted by $\mathcal{F}_x$, where $x$ is the lowest value for which at least 25 coefficient choices satisfy the criterion $\Xiexp<x$. $\mathcal{F}_2$ and $\mathcal{F}_5$ are also provided. All dimensionful quantities are given in GeV. Collider observables are the estimated number of events at an extremely hypothetical 100 TeV muon collider with 10$^{-1}$ ab integrated luminosity. Branching fractions and conversion rates are calculated assuming that the scale is set by $\Lambda_\textrm{exp}$ for the most constraining observable (usually $\mu \rightarrow e \gamma$) for the Dirac and Majorana cases, and by the scale implied by neutrino masses for the type-I seesaw case. All results are given assuming the cosmological bound on neutrino masses. 

Correlation plots for the top 100 textures of each neutrino mass model are included in the files:
\begin{itemize}[itemsep=0pt]
    \item \texttt{dirac\_correlations\_cosmo.pdf} 
    \item \texttt{majorana\_correlations\_cosmo.pdf} 
\end{itemize}
These plots show the values of observables for the top 500 trials (smallest $\Xiexp$) for each texture, smoothed using a Gaussian kernel estimation to determine approximate $1\sigma$ (solid line) and $2\sigma$ (dashed line) predictions. Shaded regions indicate parameter space that may be probed in future experiments and solid horizontal/vertical lines indicate current constraints. The constraint on the muon conversion rate in aluminum is estimated based on current constraints in gold. Collider observables are as above. The scale is set as above. 

Correlation plots for the seesaw model are not included, as nearly all textures fall outside experimental observation. 




\bibliographystyle{utphys3}
\bibliography{biblio}

\end{document}